\documentclass[fleqn,usenatbib]{mnras}

\usepackage[T1]{fontenc}
\usepackage{ae,aecompl}

\usepackage{graphicx}	% Including Figure files
\usepackage{amsmath}	% Advanced maths commands
\usepackage{amssymb}	% Extra maths symbols
\usepackage{xcolor} %package for highlighting
\usepackage{soul} %package for highlighting

\usepackage{newtxtext,newtxmath} % package moved to avoid issues in preamble

\usepackage{subcaption}
\title[Robust non-parametric morphologies]{Towards robust determination of non-parametric morphologies in marginal astronomical data: resolving uncertainties with cosmological hydrodynamical simulations}

\author[M. D. Thorp et al.]{
Mallory D. Thorp$^{1}$\thanks{E-mail: mallorythorp@uvic.ca}, Asa F. L. Bluck$^{2,3}$, Sara L. Ellison$^{1}$, Roberto Maiolino$^{2,3}$, \newauthor Christopher J. Conselice$^{4}$, Maan H. Hani$^{5,1}$, Connor Bottrell$^{6,1}$
\\
$^{1}$Department of Physics \& Astronomy, University of Victoria, Finnerty Road, Victoria, British Columbia, V8P 1A1, Canada\\
$^{2}$Kavli Institute for Cosmology, University of Cambridge, Madingley Road, Cambridge, CB3 0HA, UK\\
$^{3}$Cavendish Laboratory - Astrophysics Group, University of Cambridge, 19 J. J. Thomson
Avenue, Cambridge CB3 0HE, United Kingdom\\
$^{4}$Jodrell Bank Centre for Astrophysics, University of Manchester, Oxford Road, Manchester UK\\
$^{5}$Department of Physics and Astronomy, McMaster University, 1280 Main St. West, Hamilton ON L8S 4M1 Canada\\
$^{6}$Kavli Institute for the Physics and Mathematics of the Universe (IPMU), The University of Tokyo, Kashiwano-ha 5-1-5, Kashiwa-shi, Chiba, Japan
}

\begin{document}
\label{firstpage}
\pagerange{\pageref{firstpage}--\pageref{lastpage}}
\maketitle

% Abstract of the paper - NB there is a 200 word limit for Letters
\begin{abstract} 
Quantitative morphologies, such as asymmetry and concentration, have long been used as an effective way to assess the distribution of galaxy starlight in large samples. Application of such quantitative indicators to other data products could provide a tool capable of capturing the 2-dimensional distribution of a range of galactic properties, such as stellar mass or star-formation rate maps. In this work, we utilize galaxies from the Illustris and IllustrisTNG simulations to assess the applicability of concentration and asymmetry indicators to the stellar mass distribution in galaxies.  Specifically, we test whether the intrinsic values of concentration and asymmetry (measured directly from the simulation stellar mass particle maps) are recovered after the application of measurement uncertainty and a point spread function (PSF).  We find that random noise has a non-negligible systematic effect on asymmetry that scales inversely with signal-to-noise, particularly at signal-to-noise less than 100. We evaluate different methods to correct for the noise contribution to asymmetry at very low signal-to-noise, where previous studies have been unable to explore due to systematics. We present algebraic corrections for noise and resolution to recover the intrinsic morphology parameters. Using Illustris as a comparison dataset, we evaluate the robustness of these fits in the presence of a different physics model, and confirm these correction methods can be applied to other datasets. Lastly, we provide estimations for the uncertainty on different correction methods at varying signal-to-noise and resolution regimes.
\end{abstract}


\section{Introduction}

The characterization of galaxies according to their morphologies dates back to the very earliest studies of the extragalactic universe.  Hubble's classification scheme \citep{Hubble1926EXTRA-GALACTICNEBULAE} and the revisions that ensued over the subsequent decades \citep[e.g.][]{DeVaucouleurs1959ClassificationGalaxies, Sersic1963InfluenceGalaxy, 1998gmc..book.....V} all recognized morphology as a fundamental property of galactic populations. From these early studies, it was already clear that galaxies are broadly organized into two populations, those characterized by spheroidal shapes, and those with a prominent disk component.  Additional features, such as bars, rings and spiral arm number,  have further added to the richness of morphological classifications over the years \citep[e.g.][]{Athanassoula2009RingsMorphology, Nair2010ASURVEY, Athanassoula2013BarProperties,Willett2013GalaxySurvey}.

The importance of morphology in galaxy evolution is supported by an abundance of observational and theoretical evidence.  For example, a galaxy's internal structure is observed to strongly correlate with its global star formation rate \citep[e.g.][]{Wuyts2011GALAXY0.1, Bluck2014BulgeSurvey, Morselli2017BulgesActivity, Bluck2019WhatStructure, Cano-Diaz2019SDSS-IVSequences, Cook2020XGASS:Sequence, Sanchez2020SpatiallyGalaxies, Leslie2020TheZ5}, as well as star formation efficiency \citep{Saintonge2012THEGALAXIES,Leroy2013MOLECULARGALAXIES,Davis2014TheGalaxies,Colombo2018TheSequence, Dey2019OverviewSurveys, Ellison2021TheGalaxies}. The role of morphology in regulating star formation is also reproduced in simulations \citep[e.g.][]{Martig2009MORPHOLOGICALRED,Gensior2020HeartSpheroids}.  Other properties found to depend on morphology include gas fraction \citep[e.g.][]{Saintonge2017XCOLDStudies} and metallicity \citep[e.g.][]{Ellison2008CluesSize}. Measuring and classifying galaxy morphologies are therefore a fundamental component of any modern galaxy survey.

Whereas early morphological measurements relied on expert visual classifications, often done by individual professional astronomers, this approach has become untenable in the era of large galaxy surveys.  The deluge of galaxy images has therefore been tackled in two principal ways.  The first is through crowd-sourcing, whereby the power of the human brain continues to be tapped, through the contributions of citizen scientists \citep{Darg2010GalaxyMorphologies, Lintott2011GalaxyGalaxies, Casteels2014GalaxyInformation, Simmons2017GalaxyCANDELS, Willett2017GalaxyImaging}. Recently, artificial intelligence is replacing humans and machine learning algorithms are increasingly being applied to the challenge of large imaging datasets, either for general morphological classification \citep{Huertas-Company2015ALEARNING, DominguezSanchez2019TransferAnother, Cheng2020OptimizingImaging, 2021arXiv210208414W} or the identification of particular galaxy types/features \citep{Bottrell2019DeepRealism, Pearson2019IdentifyingLearning, 2020ApJ...895..115F,Bickley2021NoTitle}. An alternative automated approach, which has been in use for several decades, is to compute some metric of the galaxy's light distribution, a technique which is readily applicable to large datasets.  Such quantitative morphology metrics (usually) yield a continuous distribution, as opposed to placing galaxies in distinct classes.

There are numerous quantitative morphology metrics that have been developed over the years.  One of the oldest methods is the definition of a characteristic radial profile of light in a given waveband \citep{Sersic1963InfluenceGalaxy}.  Modern fitting applications of the Sérsic profile often decompose the galaxy into two (or more) components, in recognition that bulges and disks, which commonly co-exist in a given galaxy, are characterized by distinct indices \citep[e.g.][]{Simard2011ASurvey, Lackner2012AstrophysicallyGalaxies, Mendel2014ASURVEY, 2019MNRAS.486..390B,Meert2015ASystematics}.  Other metrics have been developed for specific applications, such as the identification of recent merger activity \citep[e.g.][]{Lotz2008GalaxyMergers,Pawlik2016ShapeStages}.  One of the most commonly used quantitative morphology systems is the `CAS' approach \citep{Conselice2003THEHISTORIES}, which computes concentration, asymmetry and smoothness \citep{Conselice2000TheGalaxies,Conselice2003A3,Lotz2004AClassification}. The non-parametric approach of CAS is particularly useful for identifying galaxy mergers, whose disturbed structures often do not conform to pre-defined parametric descriptions. Together, CAS and other non-parametric indices capture information that can be used to broadly distinguish early and late type galaxies, as well as possible mergers \citep{Conselice2003A3,Lotz2004AClassification,Lotz2008GalaxyMergers, 2009MNRAS.394.1956C}.

All of these quantitative morphology metrics and classifications were initially envisioned to work on flux maps originating from galactic starlight. Although some work has extended the application of traditional non-parametric indices (such as asymmetry) to the distribution of galactic gas mass \citep[e.g.][]{Holwerda2011QuantifiedGalaxies,Giese2016Non-parametricLopsidedness,Reynolds2020HGalaxies}, and stellar mass from Hubble photometry \citep{Lanyon-Foster2012TheZ1}, there has been little application to date on maps of other galactic properties, particularly the variety of properties provided by resolved spectroscopy. The widespread applicability of non-parametric morphology indicators is particularly pertinent in the era of large integral field unit (IFU) surveys wherein maps of a myriad of properties are available, i.e. the Mapping Nearby Galaxies at Apache Point Observatory (MaNGA) Survey \citep{Bundy2015OverviewObservatory,Law2015ObservingSurvey}, the Sydney-Australian-Astronomical-Observatory Multi-object Integral-Field Spectrograph (SAMI) Survey \citep{Allen2015TheRelease}, and the Calar Alto Legacy Integral Field Area (CALIFA) Survey \citep{Sanchez2012CALIFASurvey}.  Indeed, the technical challenge of effectively capturing the complexity of the spatial properties for large numbers of galaxies often leads to the use of averaged profiles, which loses much of the valuable information encoded in the IFU data \citep[e.g.][]{Ellison2018StarMaNGA,Thorp2019SpatiallyMaNGA}.  There is a clear need to explore the application of traditional techniques, and potentially develop new ones, that are capable of capturing in an effective, statistical way, the high order structure in IFU data product maps \citep[e.g.][]{Bloom2017TheFormation}.

Most importantly, the biases in non-parametric morphology measurements created by different resolutions and signal-to-noise ratio values, have not been investigated for IFU data product maps. These biases have been investigated in detail for photometry. \cite{Conselice2000TheGalaxies} demonstrated that adding simulated random noise to a galaxy image will systematically increase the measured asymmetry, and that the change in asymmetry is inversely proportional to the signal-to-noise ratio. On the same galaxy sample they also investigate the effects of degraded resolution on asymmetry, by simulating their real galaxy images as if they are observed at a larger distance, and find that asymmetry artificially decreases with distance. Degraded resolution can also alter the concentration measurement, as \cite{Bershady2000StructuralSample} demonstrated by artificially degrading their spatial sampling. The most commonly used concentration and asymmetry measurements have been constructed to mitigate these effects, and reasonable cuts in signal-to-noise and resolution can be made to make accurate asymmetry and concentration measurements within acceptable error \citep{Bershady2000StructuralSample, Conselice2000TheGalaxies, Conselice2003A3}. However, the same investigation has not been completed to see how these measurements and the necessary signal-to-noise/resolution limitations would work beyond photometric data, such as the IFU data products.

In the following work we use cosmological hydrodynamical simulations  to explore the applicability of the most commonly used non-parametric indices (asymmetry and concentration) on the kind of products available in IFU surveys. We focus our experiment on simulated maps of stellar mass, although the lessons learned from this test case can be generalized to many IFU data products.  

The paper is organized as follows.  In Section \ref{sec:methods} we describe the simulated galaxy data used in our analysis (Sec. \ref{sec:TNG}), as well as reviewing the definitions of concentration and asymmetry used in this paper (Sec. \ref{sec:CAS}).  In Section \ref{sec:analysis} we investigate how well the intrinsic concentration and asymmetry indices are recovered from the simulated galaxies once noise and point spread function (PSF) are included.   We discuss the implications of our results in Section \ref{sec:discussion} and summarize in Section \ref{sec:summary}.

\section{Data \& Methods}
\label{sec:methods}

\subsection{Simulated galaxy images}
\label{sec:TNG}

Our goal is to assess the robustness of non-parametric morphology indicators on data products other than the standard application to starlight (e.g., stellar mass maps, SFR maps, metallicity maps).  Although there is no mathematical reason that such indicators could not be applied to any 2-dimensional distribution, the limitations of realistic data may impose practical impediments.  For this reason, it has become a common approach to add observational realism to simulated data, in order to fairly compare derived properties \citep{Bottrell2017GalaxiesMass,Bottrell2017GalaxiesBiases.,Huertas-Company2019TheLearning,Bottrell2019DeepRealism,Zanisi2020TheView,2020ApJ...895..115F}.  Our approach is therefore to use simulations of galaxies, for which we can measure `true' values of a given morphology metric, before adding noise (e.g. in the form of measurement uncertainty) and the effects of degrading resolution, and comparing to the idealized measurements.

Specifically, in this work we make use of the IllustrisTNG (hereafter TNG) simulation \citep{Weinberger2017SimulatingFeedback, Pillepich2018SimulatingModel}, as well as its predecessor, Illustris \citep{2014Natur.509..177V,Genel2014IntroducingTime,Sijacki2015TheTime}, in particular we use TNG100-1 and Illustris-1 which have comparable volumes and resolutions.  The motivation for using two different simulations is to assess whether our results are sensitive to the details of the physics model used, or whether they are generalizable to different simulations or observational datasets.  The use of Illustris and TNG is ideal in this regard, since they both represent a suite of magneto-hydrodynamic (just hydrodynamic, in the case of Illustris) cosmological simulations \citep{Marinacci2018FirstFields,Naiman2018FirstEuropium,Nelson2018FirstBimodality,Pillepich2018FirstGalaxies,Springel2018FirstClustering,Nelson2019TheRelease} run with the moving-mesh code AREPO \citep{Springel2010EMesh,Pakmor2011MagnetohydrodynamicsGrid,Pakmor2016ImprovingAREPO}.  

Both simulations have runs with similar volumes and resolutions, but which employ different physical models, such as the implementation of AGN feedback \citep{Weinberger2017SimulatingFeedback}. The AGN feedback model used in TNG allows for more efficient quenching of high mass galaxies, leading to dissipationless processes that alter the morphology of quenched galaxies \citep{Rodriguez-Gomez2017TheMorphology}. Additionally, the reparameterization of galactic winds in TNG \citep{Pillepich2018SimulatingModel} results in more accurate sizes of low-mass galaxies than in Illustris. These two changes to the physical model have resulted in more realistic morphologies of TNG galaxies, as opposed to Illustris, when compared to observational data sets (e.g. \citealt{Bottrell2017GalaxiesBiases.} compared with \citealt{Rodriguez-Gomez2019TheObservations}). Thus, for the work presented here, we use TNG100-1 as our fiducial simulation, the highest resolution run of the $(110.7 \textrm{Mpc})^3$ volume, with a baryonic mass resolution of order $\sim10^6 \textrm{M}_{\odot}$. We then use Illustris-1 (hereafter Illustris), which has similar volume and resolution to TNG100-1, as a comparison sample to test the universality of the results determined from TNG.

One thousand galaxies are randomly selected each from Illustris and TNG, all of which are drawn from the final $z=0$ snapshots and have stellar masses $\textrm{M}_{\star}/\textrm{M}_{\odot}\geq10^9$. Any galaxies with $\textrm{M}_{\star}<10^9\textrm{M}_{\odot}$ are not reliably resolved. Given our random selection the $\textrm{M}_{\star}$ distribution is biased towards low mass galaxies, but as we will demonstrate in Section \ref{sec:analysis} we recover a realistic range of asymmetry and concentration values \citep{Rodriguez-Gomez2019TheObservations}. Maps of the stellar mass distribution are generated from the simulation particle data, with a gravitational softening length for stars of 0.7 kpc at z=0 \citep{Nelson2018FirstBimodality}. This idealized stellar mass map allows us to measure non-parametric morphologies without having to consider sensitivities to the mass-to-light ratio that would lead to uncertainties for an observed galaxy. Each simulated galaxy is projected along a random axis, and we select a FOV ten times the stellar half mass radius of the galaxy ($R_{\rm half}$ from here forward), with 512 pixels on each size to achieve high spatial resolution for every galaxy (compared to the resolution degradation we will explore in Section \ref{sec:psf}. 

\subsubsection{Adding Observational Effects}

Having generated the idealized stellar mass maps, observational realism is added in two steps. First, we consider the effect of degrading the resolution of the stellar mass map, which will reduce the contrast of asymmetry structures and alter the concentration measurement \citep{Conselice2000TheGalaxies,Bershady2000StructuralSample}. We achieve this by convolving the stellar mass map with a Gaussian PSF that has a full width at half maximum (FWHM) ranging between 0.002"-7" (with the goal of ranging from a negligible PSF to some of the largest PSFs in current observational surveys). We apply an arcsecond PSF value by assuming each galaxy is at the average distance of a MaNGA survey galaxy z=0.037 (the applicable distance for comparing this analysis to IFU data), and assuming a cosmology in which $\mathrm{H}_0=$ 70 km $\mathrm{s}^{-1}$ $\mathrm{Mpc}^{-1}$, $\Omega_{\mathrm{m}}=$0.3, and $\Omega_{\Lambda}=$0.7.

\begin{figure*}
	\includegraphics[width=0.9\textwidth]{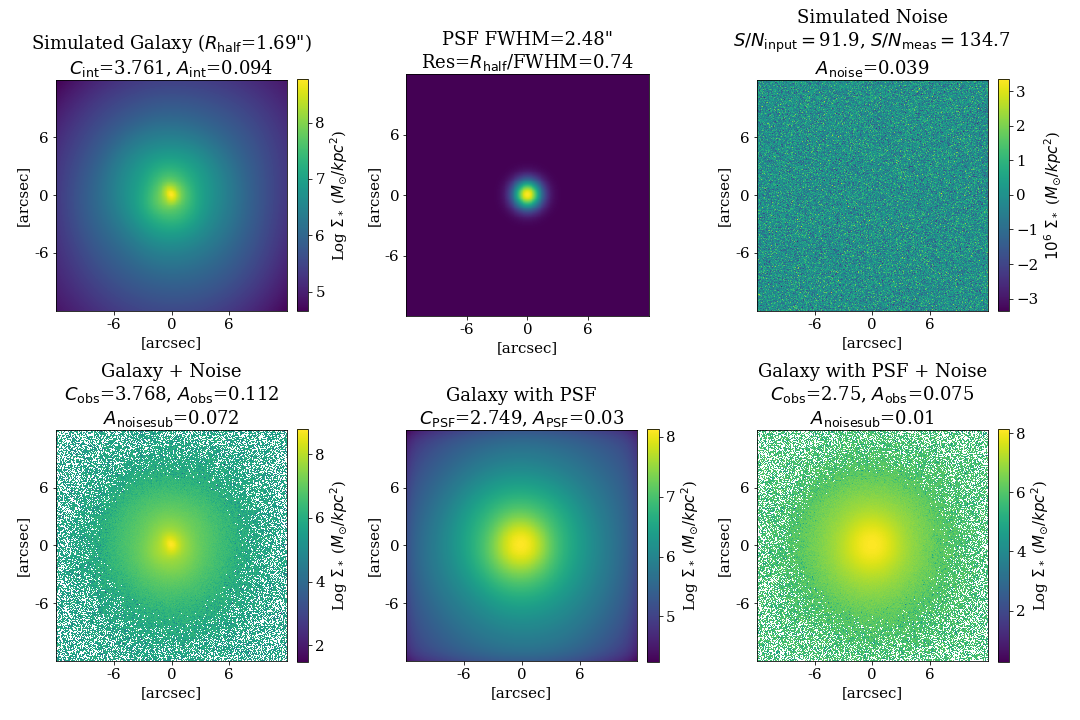}
	\centering
    \caption{The steps through which we apply realistic noise and resolution to a simulated galaxy. Top-left: The stellar mass surface density of the simulated galaxy, from which intrinsic asymmetry and concentration is measured. Top-middle: A sample PSF applied to this galaxy, though values can range from 0.002"-7". Changing the PSF allows us to change the resolution to our simulated galaxies at a fixed distance. We chose to highlight this galaxy with a small angular size compare to the applied PSF, resulting in a small resolution value (smaller than has been examined in previous works), to highlight how significantly degrading the resolution changes the asymmetry and concentration. Top-right: Random noise generated from a Gaussian distribution, to be applied to the galaxy image. The desired (input) $S/N$ is used to generate the width of the Gaussian distribution so that the measured $S/N$ is of a similar (though not exactly the same) value. Bottom-left: The stellar mass surface density image of the galaxy with added noise. White values are where $\Sigma_{\star}$ is so small, adding negative noise results in a non-measurement. Bottom-middle: The image of the galaxy convolved with the PSF from the top-middle panel. Bottom-right: The image of the galaxy convolved with the PSF and with noise added, providing an image of the galaxy more similar to what would be seen with observations.}
    \label{fig:Ex_Noise_PSF}
\end{figure*}

The PSF is only a component of the more fundamental resolution of the image, which itself is the relationship between the convolving PSF size and the apparent size of the galaxy. The set physical size of the galaxy, along with our choice of a constant distance, means that the apparent angular size of the galaxy is fixed and one need only vary the PSF to vary the resolution. Resolution ("Res") is defined in this analysis to be the ratio of the stellar half mass radius ($R_{\mathrm{half}}$) in arcseconds and the PSF FWHM in arcseconds (Res=$R_{\mathrm{half}}$/FWHM). By using $R_{\mathrm{half}}$ there is no need to fit the galaxy structure to measure radius, though we note that any approximation of galaxy size with respect to PSF would suffice. Figure \ref{fig:Ex_Noise_PSF} summarizes the different levels of realism applied to an example TNG galaxy stellar mass map (top left panel). The top middle panel shows an example PSF similar to that in the MaNGA survey (2.48"), but given the small angular size of the galaxy ($R_{\mathrm{half}}$=1.69") the resolution of the image is actually very low.

Second, noise is added to the mass map.  As we explain in the next sub-section, the traditional asymmetry index involves the subtraction of a background sky term, in order to account for observational effects that contribute to asymmetry, but which is not associated with the galaxy itself.  This is akin to subtracting the sky from a photometric measurement.  As we will show in the next sub-section, the subtraction of the background sky term  has a significant effect on the traditional asymmetry index, due to its mathematical definition. Although we are working with stellar mass maps, rather than light, noise would still present in an observed stellar mass map in the form of measurement and post-processing uncertainties. We therefore create a random `noise' map to add to the simulated galaxy mass map, to approximate an observational scenario where every pixel has some hypothetical uncertainty on the mass measurement. From this noise map we can measure a signal-to-noise to quantify the significance of the contribution from uncertainty, with the goal of creating a variety of noise maps to span a large range of possible signal-to-noise values.

To generate a random noise map representing the hypothetical uncertainties on each pixel for the simulated stellar mass map, we first select a desired signal-to-noise ratio ranging from 0 to 1000. We then measure the median signal of the galaxy within a $1R_{\rm half}$ aperture (where $R_{\rm half}$ is the stellar half-mass radius). We approximate the median absolute deviation (MAD) within that aperture to be the median signal within that aperture divided by the desired signal-to-noise ratio. Given we set the noise map to be a normal distribution, we can convert the MAD to a standard deviation by multiplying by the known scale factor $k=1/(\Phi^{-1}(3/4))=1.4826$ (where $\Phi^{-1}$ is the inverse of the cumulative distribution, and $3/4$ implies $\pm$MAD covers half the standard normal cumulative distribution function). Using the the standard deviation we construct a standard-normal distribution centred on zero; for each pixel we select a random value from this distribution to set the uncertainty of that pixel. This method ensures that the resulting signal-to-noise (the median value of the mass map within $1R_{\rm half}$ divided by the median value of the noise map within $1R_{\rm half}$) is approximately (though not exactly) equal to the desired signal-to-noise used to construct the noise map.

An example noise field is shown in the top right panel of Fig. \ref{fig:Ex_Noise_PSF}. The lower panels of Figure \ref{fig:Ex_Noise_PSF} show the resulting mass maps when either the noise (left panel), the PSF (middle panel) or both (right panel) are included.

\subsection{Nonparametric morphology indicators}
\label{sec:CAS}

Each of the 1000 TNG galaxies is assessed using two non-parametric indices: concentration and asymmetry, both of which are computed using the \texttt{statmorph} package \citep{Rodriguez-Gomez2019TheObservations}\footnote{https://statmorph.readthedocs.io/en/latest/overview.html}.  We refer the reader to the original papers for details of the development of these metrics \citep{Schade1995Canada-FranceGalaxies,Abraham1996GalaxyField,Bershady2000StructuralSample,Conselice2000TheGalaxies,Conselice2003A3}, but review the basic details here.  

The concentration parameter (C) has been used for over half a century \citep[see][]{Morgan1959AIL}.  Although there is a plethora of possible definitions of concentration, the most commonly used in the assessment of galaxy light distributions is that of \cite{Bershady2000StructuralSample} and \cite{Conselice2003THEHISTORIES}.  These works define concentration, as:

\begin{equation}
    C = 5 \log_{10} (r_{80}/r_{20})
\end{equation}

\noindent where $r_{80}$ and $r_{20}$ are the radii within which 80\% and 20\% of the galaxy's total flux (for this work, total stellar mass) is contained, respectively. The inner and outer radii are chosen specifically to mitigate the degradation in concentration that comes from low resolution (see \cite{Bershady2000StructuralSample} for details).  In the work presented here, we compute the intrinsic concentration ($C_{\rm int}$) derived directly from the idealized simulation image. Once we degrade the resolution of the image by convolving with a PSF, we refer to the measured concentration as $C_{\rm Res}$; comparing this value to $C_{\rm int}$ informs us of the impact of resolution on the true concentration.  Finally, we define $C_{\rm obs}$ as the measured concentration once noise has been added to the image.  Again, comparison of $C_{\rm obs}$ to $C_{\rm int}$ will quantify how much the intrinsic concentration of the simulated galaxy has been affected by realistic observational features.

Identifying the galaxy's centre is critical to the calculation of concentration, as well as many other non-parametric indices, including the second index of interest in this work asymmetry. Therefore concentration and asymmetry are often computed in conjunction, in an iterative approach that finds a centre which yields minimum asymmetry value, then uses that centre to measured concentration as well. As the name implies, the asymmetry parameter quantifies how much of an image's light is not symmetric about an axis of rotation, calculated as the difference between an image and its 180\textdegree rotation \citep{Conselice2000TheGalaxies}.

The intrinsic asymmetry ($A_{\rm int}$) is defined as:

\begin{equation}
    A_{\rm int} \equiv \frac{\Sigma_{i,j} \mid I_{ij} - I^{180}_{ij}\mid} {\Sigma_{i,j}\mid I_{ij}\mid}
    \label{eqn:A_int}
\end{equation}

\noindent where $I_{i,j}$ is the flux contained within an individual pixel of the galaxy image, and $I^{180}_{i,j}$ is the pixel at that same $i,j$ location once the image is rotated 180 degrees. These values are computed within 1.5 times the Petrosian radius ($r_{\rm petro}$).  For an idealized image, such as one generated from a simulation, $A_{\rm int}$ is a perfect representation of the galaxy's asymmetry (with no background noise).

However, in real images there are observational artefacts that can contribute to the asymmetry that are not intrinsic to the galaxy, such as sky noise and background gradients introduced by nearby bright sources.   For either a real galaxy observation, or a simulated galaxy that has had noise added to it, the measured asymmetry is therefore a combination of the galaxy signal and the background noise:

\begin{equation}
    A_{\rm obs} = \frac{\Sigma_{i,j} \mid (I+\mathrm{noise})_{ij} - (I+\mathrm{noise})^{180}_{ij}\mid} {\Sigma_{i,j}\mid (I+\mathrm{noise})_{ij}\mid}
    \label{eqn:A_obs}
\end{equation}

Therefore, in order to recover the intrinsic asymmetry, the noise asymmetry ($A_{\rm noise}$) needs to be quantified and removed from the observed asymmetry.  This is analogous to the subtraction of background sky in the measurement to obtain galaxy photometry.  In observations, $A_{\rm noise}$ is computed from a region offset from the galaxy; in simulations it can alternatively be computed from the noise (background) field generated in addition to the idealized image, i.e.:

\begin{equation}
    A_{\rm noise} = \frac{\Sigma_{i,j} \mid \mathrm{noise}_{ij} - \mathrm{noise}^{180}_{ij}\mid} {\Sigma_{i,j}\mid \mathrm{noise}_{ij}\mid}
    \label{eqn:A_sky}
\end{equation}

The premise above is that by subtracting the contribution to the measured asymmetry ($A_{\rm obs}$) from the background noise ($A_{\rm noise}$, a term that captures all background sources of asymmetry that we wish to remove), we can approximate the intrinsic asymmetry.  That is, we determine a corrected, noise-subtracted asymmetry $A_{\rm noisesub}$ as: 

\begin{equation}
    A_{\rm noisesub} \equiv A_{\rm obs} - A_{\rm noise}
\end{equation}

\noindent with the assumption that $A_{\rm noisesub} \sim A_{\rm int}$.

However, due to the modulus present in the equations for $A_{\rm obs}$, $A_{\rm noise}$, and $A_{\rm int}$, it is not mathematically correct to assume that $A_{\rm int}$=$A_{\rm noisesub}$. Given the rule of modular subtraction ($\mid A\mid - \mid B\mid \leq \mid A-B\mid$),  $A_{\rm noisesub}$ can only serve as a lower limit to the true value of $A_{\rm int}$.  Indeed, due to the presence of the modulus in Eqn \ref{eqn:A_sky} even a random noise field will have a non-zero asymmetry associated with it. The underestimation of asymmetry through this noise correction method had been previously investigated in photometric studies, which recommend the asymmetry measurement only be made where $S/N>100$. Otherwise noise dominates the asymmetry, see \citealt{Conselice2000TheGalaxies}. Whereas this regime is readily achievable in photometry, such a high $S/N$ is unlikely to be reached  in maps of galactic properties, and previous works on the asymmetry of galactic gas and dust have struggled with this as well \citep{Bendo2007VariationsSequence,Giese2016Non-parametricLopsidedness,Reynolds2020HGalaxies}. Although we are working with stellar mass maps, which don't suffer from a sky contribution, noise is still present in observed stellar mass maps in the form of uncertainties in the determination of the stellar mass itself, and these uncertainties are expected to far exceed the equivalent of a $S/N$ $\sim$ 100 criterion \citep[e.g.][]{Mendel2014ASURVEY, Sanchez2016Pipe3DFIT3D,Sanchez2016Pipe3DDataproducts}.

Figure \ref{fig:Ex_Noise_PSF} shows how asymmetry and concentration can be affected by resolution and noise, highlighting the impact of noise subtraction on the recovered asymmetry. In the example galaxy shown in Figure \ref{fig:Ex_Noise_PSF} the intrinsic concentration is measured to be $C_{\rm int}$=3.761 (top left panel). After degrading the resolution with Res=0.74 (the stellar half mass radius is smaller than the PSF that convolves the image), the concentration is significantly diminished to $C_{\rm obs}$=2.749 (lower middle panel).  Conversely, if just noise is added to the image (no degraded resolution), the measured concentration is essentially unaffected, with a difference less than 0.1\% from the intrinsic value.  The combined effect of resolution and noise yields an observed concentration of $C_{\rm obs}$=2.75, i.e. a net decrease from the true value by $\sim$27 per cent.  As we will show for the full galaxy sample in the next section, observational changes in concentration can be traced solely to the value of resolution.

As for the impact of resolution and noise on asymmetry, the example galaxy in Figure \ref{fig:Ex_Noise_PSF} starts with an intrinsic asymmetry $A_{\rm int}$=0.094 (top left panel).  The particular noise field generated for this example has an asymmetry (introduced by the use of the modulus in Equation \ref{eqn:A_sky}) of $A_{\rm noise}$=0.039.  Since the measured asymmetry in the noisy image for this example galaxy is $A_{\rm obs}$=0.112 (lower right panel), subtracting the noise asymmetry yields a corrected value of $A_{\rm noisesub}=0.073$. The noise-subtracted asymmetry is therefore, in this case, under-estimating the true asymmetry by a factor of  $\sim$20\%. We next turn to examine the effects of a broad range of resolution and signal-to-noise values.  In the following section, we compare the measured, corrected and intrinsic morphologies of our full simulated galaxy sample, and investigate whether further corrections could be made to improve them.

\section{Analysis}
\label{sec:analysis}

In this section we assess the concentration and asymmetry of our sample of 1000 galaxies in the TNG simulation (a comparison with the Illustris simulation is given in Section \ref{sec:compare}).  We assess separately the impacts of noise (which in the case of observed mass maps is driven by uncertainty on the measurements) and resolution, quantifying how well the intrinsic parameters can be recovered once these compounding observational effects are included.  Finally, we investigate whether improvements can be made to the traditional definitions of concentration and asymmetry that will increase the fidelity of these metrics in marginal observational data.

\subsection{Noise}
\label{sec:noise}

\begin{figure*}
	\includegraphics[width=0.9\textwidth]{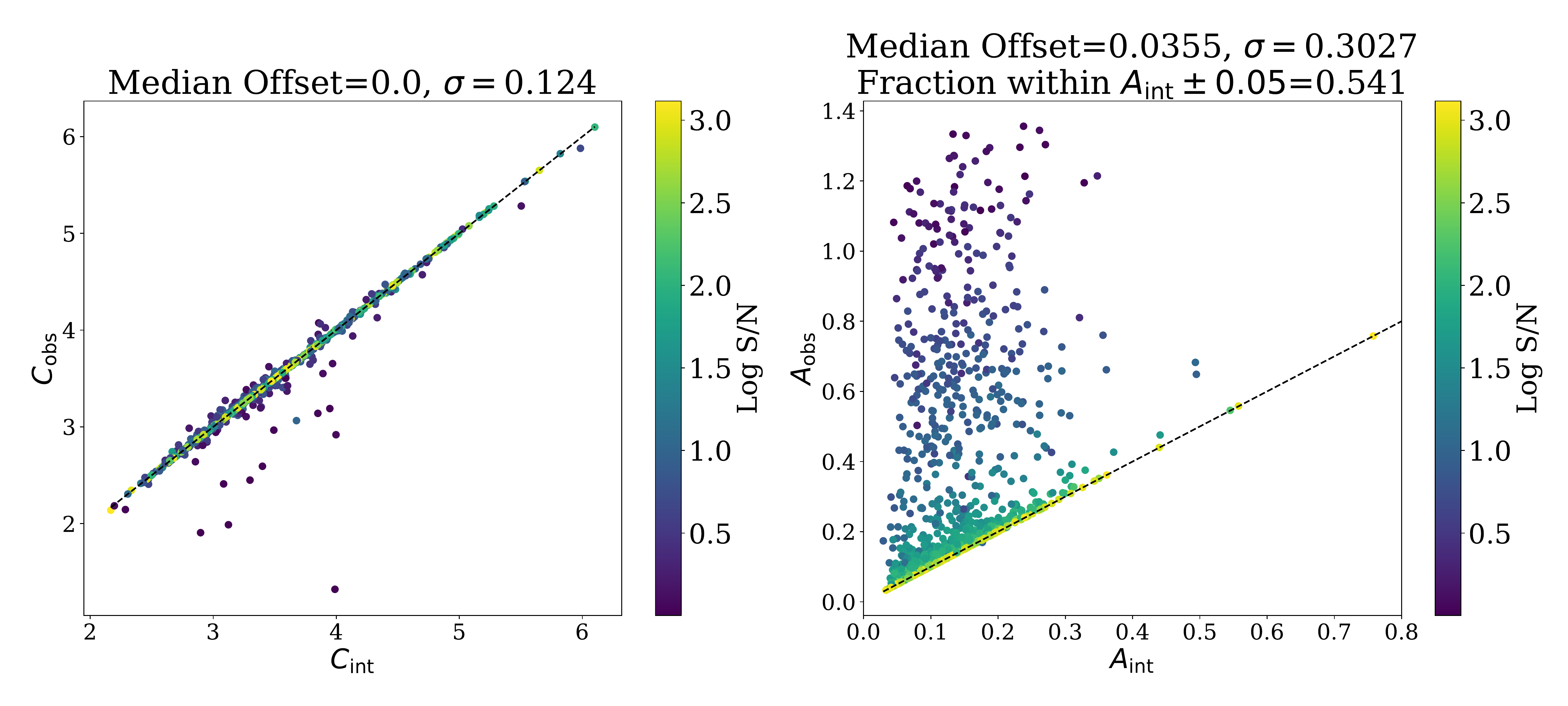}
	\centering
    \caption{The observed concentration (left) and asymmetry (right) compared to intrinsic values (which have no random noise contribution). From the left panel it is clear that the addition of noise has no effect on concentration measurements (save catastrophic cases at the lowest signal-to-noise). Asymmetry, on the other had, is always overestimated once noise is added, though the degree of that overestimation scales with inversely with $S/N$.}
    \label{fig:Int_Noise}
\end{figure*}

Figure \ref{fig:Int_Noise} demonstrates how adding random noise to the stellar mass map changes the observed asymmetry and concentration in the 1000 galaxies taken from the TNG simulation. Noise contribution has little impact on the observed concentration (left panel).  This result is expected, since a randomly constructed noise field will contribute uniformly to the inner and outer apertures from which concentration is calculated. The only exception to this rule is galaxies with the lowest signal-to-noise ($S/N<5$), where random clumping of pixels with high noise can shift the concentration measurement (hence why the scatter is also unbiased). Thus, when measuring the concentration of an image, one generally need not worry about any intrinsic bias created by the noise.

In contrast, adding noise to the image always leads to an over-estimate of the observed asymmetry (Equation \ref{eqn:A_obs}) compared with its intrinsic value (right panel of Figure \ref{fig:Int_Noise}), though the change is greatest for galaxies with low signal-to-noise, when the noise dominates the image \citep[see][]{Conselice2000TheGalaxies,Lotz2004AClassification,Giese2016Non-parametricLopsidedness}. For $S/N>100$ the difference rarely exceeds a 50$\%$ increase, but for $S/N<100$ the asymmetry can increase anywhere between a factor of 2 to 10. To reiterate, because of the modulus within the asymmetry measurement even a random noise map will have a net-positive value. 

\begin{figure}
	\includegraphics[width=0.9\columnwidth]{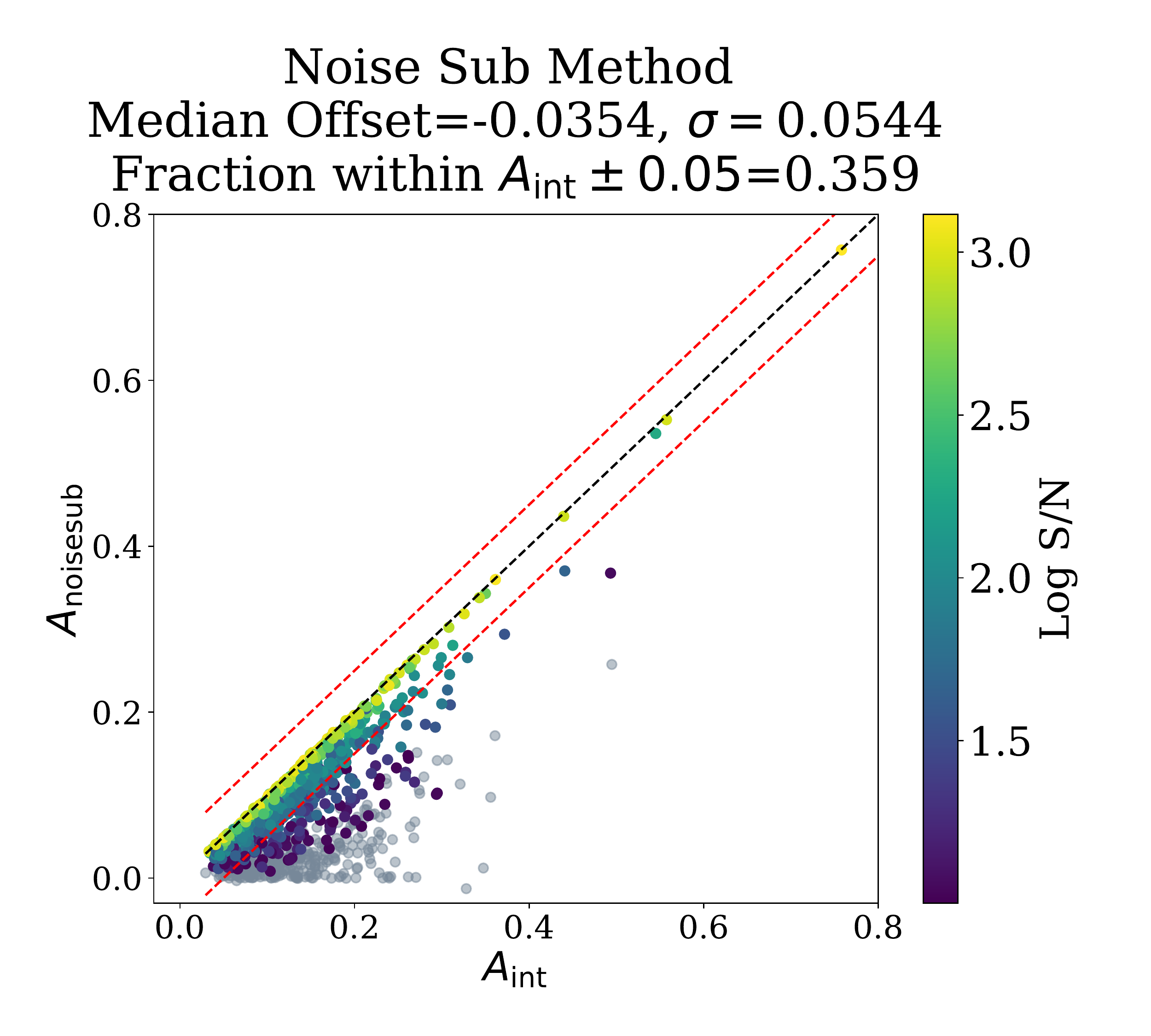}
	\centering
    \caption{The standard noise subtraction method requires subtracting all of $A_{\rm noise}$ from $A_{\rm obs}$ to approximate the intrinsic asymmetry. The black dashed line represents the line of equality, with red dashed lines representing a 0.05 buffer from this line. Points are colour-coded by $S/N$, save for galaxies with $S/N<10$, which are grey. By correcting for noise contribution to asymmetry the scatter between the measured and intrinsic asymmetry is lessened by a factor of 10, and ~5\% more galaxies have a good asymmetry measurement (within 0.05 of $A_{\rm int}$). However, the median offset using this correction is of a similar magnitude to doing no noise correction (see Figure \ref{fig:Int_Noise}), just in the opposite direction. The galaxies with asymmetries far less than $A_{\rm int}\pm0.05$ are those with low $S/N$, in particular the galaxies with $S/N<10$ have the greatest difference in asymmetry from noise.}
    \label{fig:Noise_Corr}
\end{figure}

The noise subtraction method has traditionally been used to account for the non-negligible addition of noise, however, as discussed in Section \ref{sec:CAS}, it only works well for the $S/N>100$ regime (when noise does not dominate the asymmetry). Figure \ref{fig:Noise_Corr} illustrates how subtracting the noise contribution to asymmetry decreases the scatter between the measured and intrinsic asymmetry. $A_{\rm noisesub}$ values for galaxies with $S/N>100$ (the $S/N$ cut suggested by \citealt{Conselice2000TheGalaxies}) are within $\pm 0.05$ of their intrinsic value (represented by the grey dashed lines). However, galaxies with $S/N<100$ now have a preferentially underestimated asymmetry (creating a similar problem, although in the opposite direction, to the overestimation that occurs with no $A_{\rm noise}$ correction).

The presence of a bias in the asymmetry correction method is obviously problematic;  despite being introduced as an improvement in the estimate of asymmetry of intermediate signal-to-noise galaxies (where background noise can have a non-negligible contribution), it actually introduces a new offset in the low $S/N$ regime. However, given $A_{\rm obs}$ is an overestimation and $A_{\rm obs} - A_{\rm noise}$ is an underestimation of the true asymmetry, one could postulate that a noise correction that more accurately reproduces $A_{\rm int}$ lies somewhere in between. We propose that subtracting a fraction of $A_{\rm noise}$, rather than the entire value, could mitigate the over-correction. By subtracting multiple different fraction of $A_{\rm noise}$ from $A_{\rm obs}$ we can determine what fraction of the noise needs to be corrected to reproduce $A_{\rm int}$.

\begin{figure*}
	\includegraphics[width=0.9\textwidth]{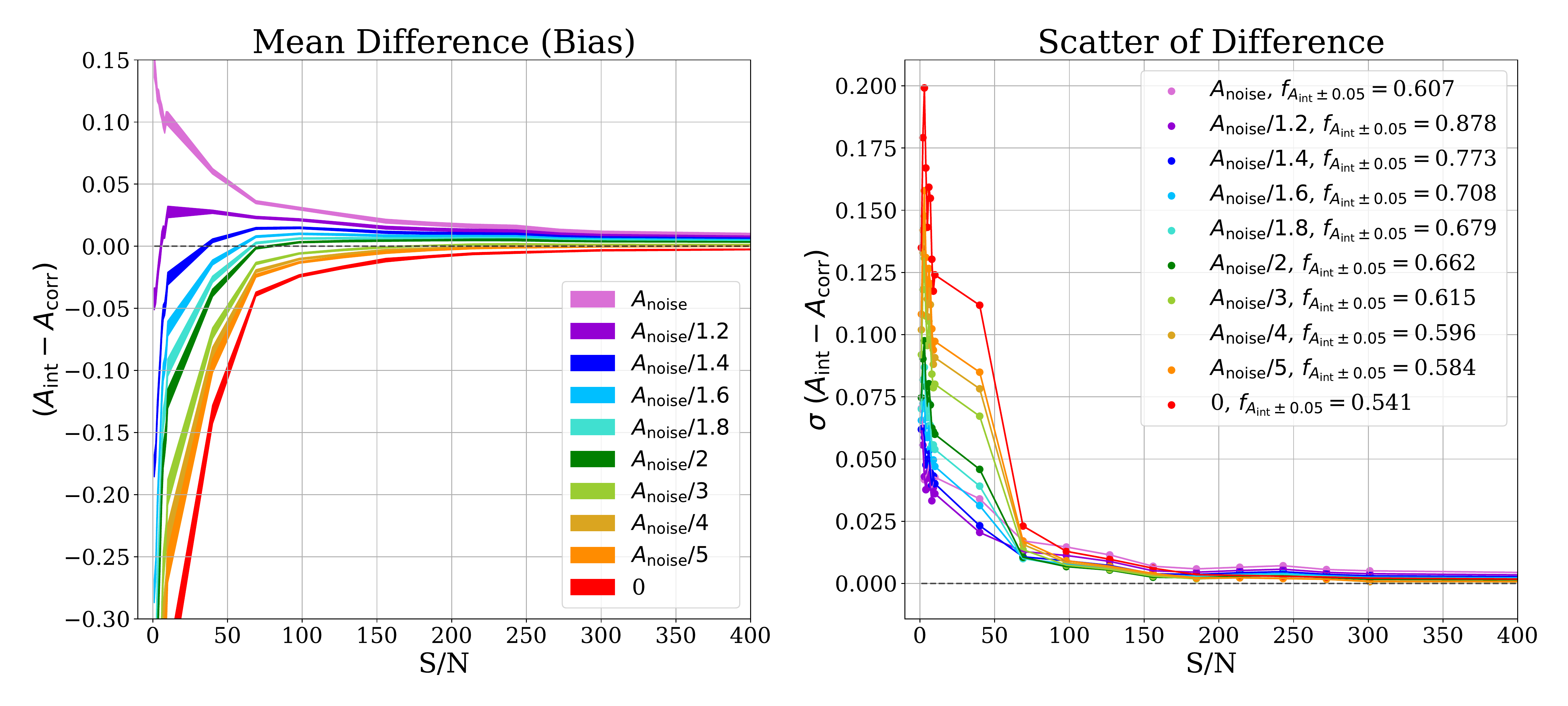}
	\centering
    \caption{Left: The mean difference between the intrinsic asymmetry and corrected asymmetry for increasing signal-to-noise. Each curve represents subtracting a different fraction of the noise asymmetry, where red has no subtraction, and purple subtracts the entire $A_{\rm noise}$ value. The width of each line is the standard error on the mean. See Figure \ref{fig:TNG_A_corr} for a comparison of these plots to our fit correction. Right: The standard deviation of the difference between intrinsic and corrected asymmetry in each signal-to-noise bin, for each correction method. From this we can see that to recover the intrinsic asymmetry at $S/N<100$, subtracting a fraction of $A_{\rm noise}$ does better than subtracting the total noise contribution to asymmetry. Subtracting $A_{\rm noise}/1.2$ recovers the most asymmetry values (within $A_{\rm int}\pm0.05$) and has the smallest scatter in the difference at all $S/N$ values.}
    \label{fig:divide_by}
\end{figure*}

\begin{figure}
	\includegraphics[width=0.9\columnwidth]{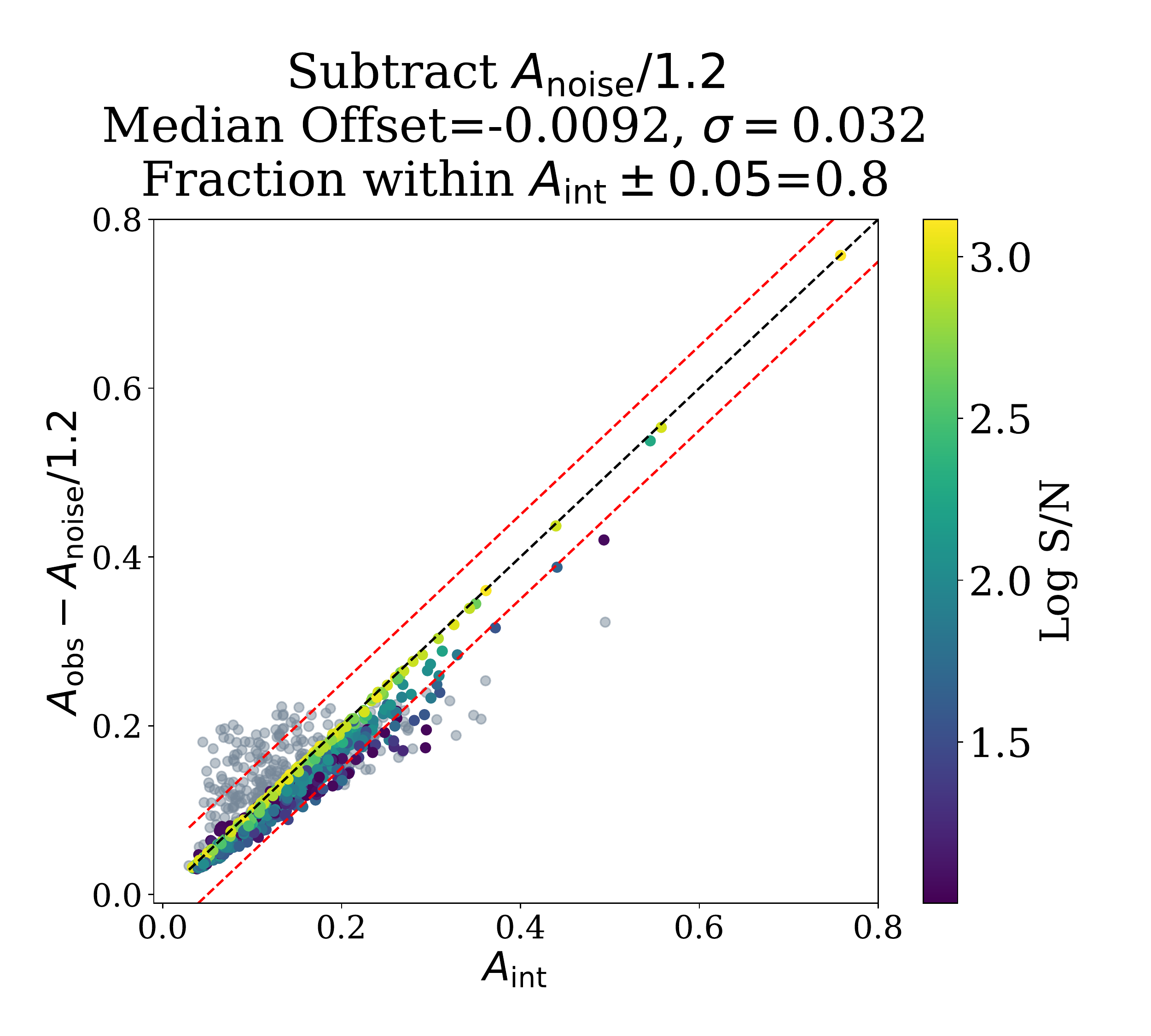}
	\centering
    \caption{The asymmetry corrected for $A_{\rm noise}$/1.2 versus the intrinsic asymmetry. Points are colour-coded by $S/N$, except for galaxies with $S/N<10$ which are grey. The scatter between the two is smaller than doing no noise correction, similar to the improvement from subtracting $A_{\rm noise}$ from the observed asymmetry (see Figure  \ref{fig:Noise_Corr}). However, the median offset is significantly less, and over 20\% more of the simulated galaxies are within an accurate asymmetry measurement (within $A_{\rm int}\pm0.05$). In fact, almost all galaxies with $S/N>10$ are within 0.05 of $A_{\rm int}$, a much more forgiving $S/N$ cut than the $S/N>100$ cut necessary for the traditional $A_{\rm noise}$ correction.}
    \label{fig:Noise_Corr_frac}
\end{figure}

Figure \ref{fig:divide_by} exhibits how the difference between the intrinsic and noise corrected asymmetry varies with $S/N$ for multiple fractions of $A_{\rm noise}$ correction. Each coloured curve in Figure \ref{fig:divide_by} represents a different fractional correction of $A_{\rm noise}$, with pink representing a full subtraction of $A_{\rm noise}$, and red representing no subtraction. Both subtracting the sky asymmetry and doing no subtraction only achieve reasonable accuracy within $A_{\rm int}\pm0.05$ for a signal-to-noise greater than 100, when the curves begin to approach zero difference between the true and observed asymmetry. However, varying the fractional value of $A_{\rm noise}$ improves the recovery of the intrinsic asymmetry to lower signal-to-noise values. The best option between doing nothing and subtracting the total $A_{\rm noise}$ is actually to divide by 1.2, which results in the greatest fraction of galaxies within 0.05 of $A_{\rm int}$ (excluding galaxies with $S/N<10$ to ensure answers aren't skewed by the asymptotic nature of each curve). correcting for $A_{\rm noise}$/1.2 also results in the smallest scatter in the difference for all $S/N$ regimes. Figure \ref{fig:Noise_Corr_frac} directly compares the measured and intrinsic asymmetry (as was completed in Figure \ref{fig:Noise_Corr} for traditional noise correction). Though the scatter in the difference between the two is only marginally better than regular $A_{\rm noise}$ correction, the median offset has diminished and over 20\% more galaxies have reliable asymmetry measurements (within 0.05 of the intrinsic asymmetry). Therefore, when considering low $S/N$ data using $A_{\rm obs} - A_{\rm noise}$/1.2 is the best way to approximate $A_{\rm int}$, and we will continue to investigate this as a candidate asymmetry measurement for the rest of the analysis (along with correcting for $A_{\rm noise}$, and doing no correction).

\subsubsection{Fit to Recover Intrinsic Asymmetry for all $S/N$}

The two methods to correct for noise asymmetry described so far are approximations which work within a signal-to-noise limited system. However, if there exists some mathematical relationship between the intrinsic asymmetry, the observed asymmetry, and the signal-to-noise, one should be able to fit a relationship between two of those parameters to almost perfectly approximate the third. In order to explore this possibility, we assume a relationship between the intrinsic asymmetry, the observed asymmetry, and the signal-to-nose of the following form:

\begin{equation}
    A_{\rm int} = (F_1(S/N)+1)A_{\rm obs} +F_2(S/N)A_{\rm obs}^2+ F_3(S/N)A_{\rm obs}^3
    \label{Eqn:A_int_blank}
\end{equation}

\noindent where $F_3$, $F_2$, and $F_1$ are each unique functions of signal-to-noise of the following form:

\begin{equation}
    F(S/N) = X_1(S/N)^{-1}+X_2(S/N)^{-2}+X_3(S/N)^{-3}
    \label{Eqn:Polynomial_SN}
\end{equation}

\noindent Equation \ref{Eqn:A_int_blank} is designed so that as $S/N$ goes to infinity, $A_{int}=A_{obs}$. As we established in Figure \ref{fig:Int_Noise}, at the highest $S/N$ the observed asymmetry is almost exactly equal to the intrinsic asymmetry, as the contribution from uncertainty is practically negligible. Thus we wish to replicate a relationship between $A_{\rm int}$, $A_{\rm obs}$, and $S/N$ that aligns with the fact that as the uncertainty goes to zero (and $S/N$ goes to infinity), $A_{\rm int}$ should equal $A_{\rm obs}$.

We elect to only fit data with $S/N<500$. At $S/N>500$, the difference between the observed and intrinsic asymmetry never exceeds 1\% of $A_{\rm int}$, and it is therefore safe to assume that $A_{\rm int} \approx A_{\rm obs}$. By excluding high $S/N$ data from the fit we can focus on best characterizing the asymptotic change of asymmetry with $S/N$. We fit Equation \ref{Eqn:A_int_blank} to the $A_{\rm obs}$ and $S/N$ of the 788 TNG galaxies with $S/N<500$, using the python package LM-Fit\footnote{https://dx.doi.org/10.5281/zenodo.11813}, which minimises the root-mean-square difference between the intrinsic asymmetry and the observed asymmetry. The following fit is found to minimise the difference between $A_{\rm int}$ and $A_{\rm obs}$:

\begin{equation}
    \begin{aligned}
    &   \textbf{S/N<500: } A_{\rm int} = (- (15.36\pm0.58)(S/N)^{-1}+ (78.85\pm3.38)(S/N)^{-2} \\ & -(86.89\pm4.41)(S/N)^{-3} +1)A_{\rm obs} + ((10.69\pm1.39)(S/N)^{-1} \\ &  - (99.59\pm6.37)(S/N)^{-2} + (125.44\pm7.63)(S/N)^{-3} )A_{\rm obs}^2 \\ &  + (-(0.91\pm0.83)(S/N)^{-1}+ (31.98\pm3.19)(S/N)^{-2} \\ & -(45.82\pm3.43)(S/N)^{-3})A_{\rm obs}^3\\
    &   \textbf{S/N>500: } A_{\rm int} = A_{\rm obs} \\
    \end{aligned}
    \label{eqn:A_noise_fit}
\end{equation}

The coefficients in Equation \ref{eqn:A_noise_fit} are specific to how we measure the signal-to-noise ratio, in particular using $R_{\rm half}$ to set the aperture from which we make that measurement. In practice one could repeat the fitting process laid out thus far using a different aperture radius (as would be necessary for different kinds of observations). This would significantly change the coefficients present in the final relationship between $A_{\rm int}$, $A_{\rm obs}$, and $S/N$. But the conceptualization of this fit, as well as its relative accuracy, would remain unchanged.

\begin{figure}
	\includegraphics[width=0.9\columnwidth]{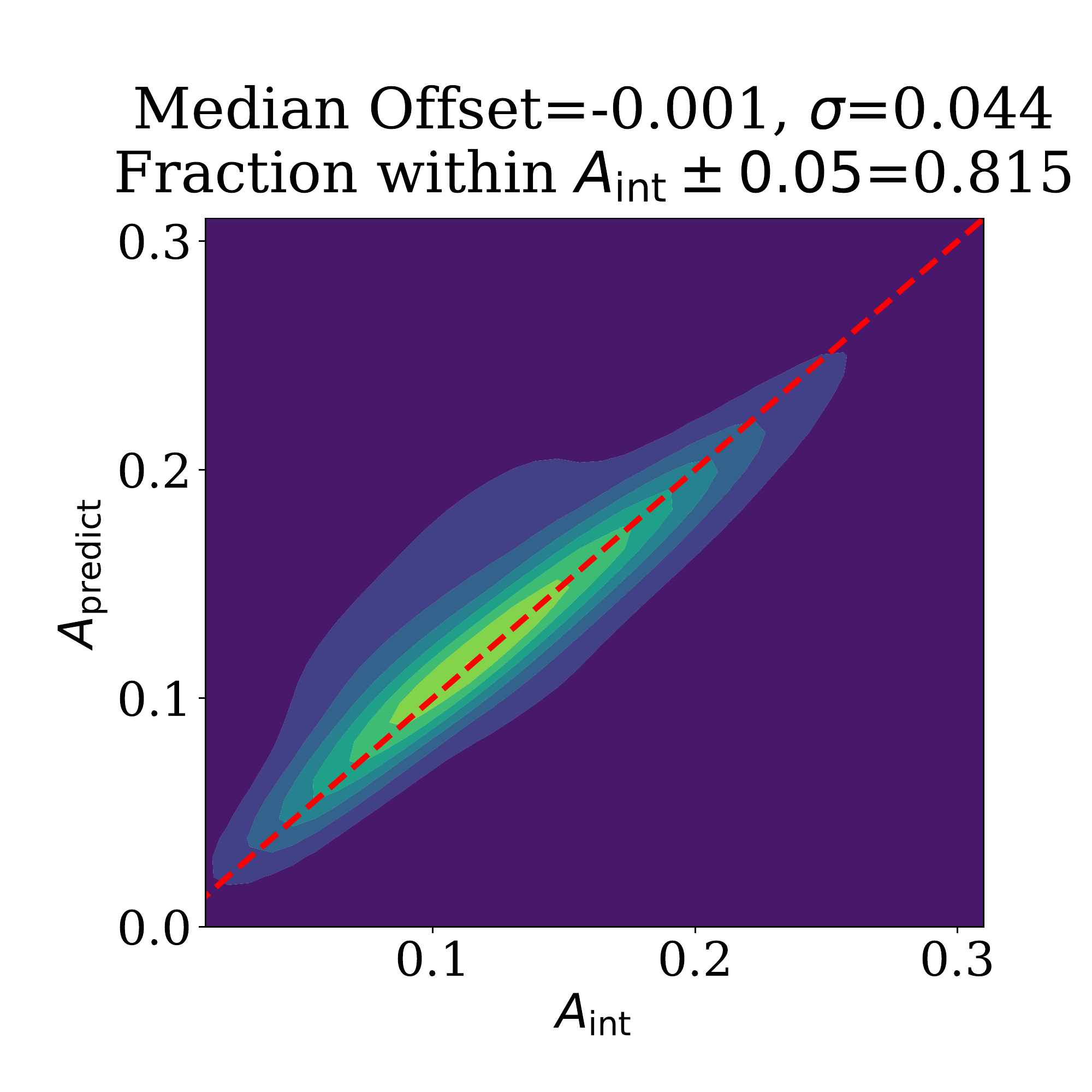}
	\centering
    \caption{Density contours for the predicted asymmetry based on the fit relationship from Equation \ref{eqn:A_noise_fit} versus the intrinsic asymmetry, the dashed red line representing where the two would be equal to each other. Applying a correction for the random noise based on a fit function significantly decreases the scatter of the difference between the predicted and intrinsic asymmetry, while also removing any bias in the predicted asymmetry. The scatter in the fit is dominated entirely by low $S/N$ galaxies.}
    \label{fig:Fit_scat}
\end{figure}

We evaluate the goodness of this fit by comparing the intrinsic asymmetry to that predicted by Equation \ref{eqn:A_noise_fit} in Figure \ref{fig:Fit_scat} (see Section \ref{sec:compare} for tests on independent samples). Both the median offset between the observed asymmetry and intrinsic asymmetry, and the scatter of this offset, is reduced by a factor of 10 after correcting asymmetry based on this fit. The majority of the scatter here comes from the lowest $S/N$ ($<10$) galaxies, where the accuracy of the fit is lowest. However, low $S/N$ galaxies can both have over- or under-estimated asymmetry measurements (unlike the bias to one or the other when correcting based on $A_{\rm noise}$). The difficulty with fitting the low signal-to-noise data likely stems from the asymptotic relationship between the intrinsic and observed asymmetry at low signal-to-noise, thus a small change in the fit can lead to a more drastically incorrect prediction. Of the predicted asymmetries, including all $S/N$ values, 81.5\% are within 0.05 of $A_{\rm int}$ (what we would consider a "good" asymmetry measurement). Thus the fit correction is a significant improvement on the 54.1\% and 60.7\% for doing no correction, and subtracting $A_{\rm noise}$, respectively. Subtraction $A_{\rm noise}$/1.2 is still the best asymmetry prediction by a margin, recovering 87.8\% of asymmetries within 0.05 of $A_{\rm int}$. Though the fit correction still has its own benefits over the $A_{\rm noise}$/1.2 correction, in particular when simultaneously correcting for resolution as is investigated in Section \ref{sec:Res_A}.

\begin{figure*}
	\includegraphics[width=0.9\textwidth]{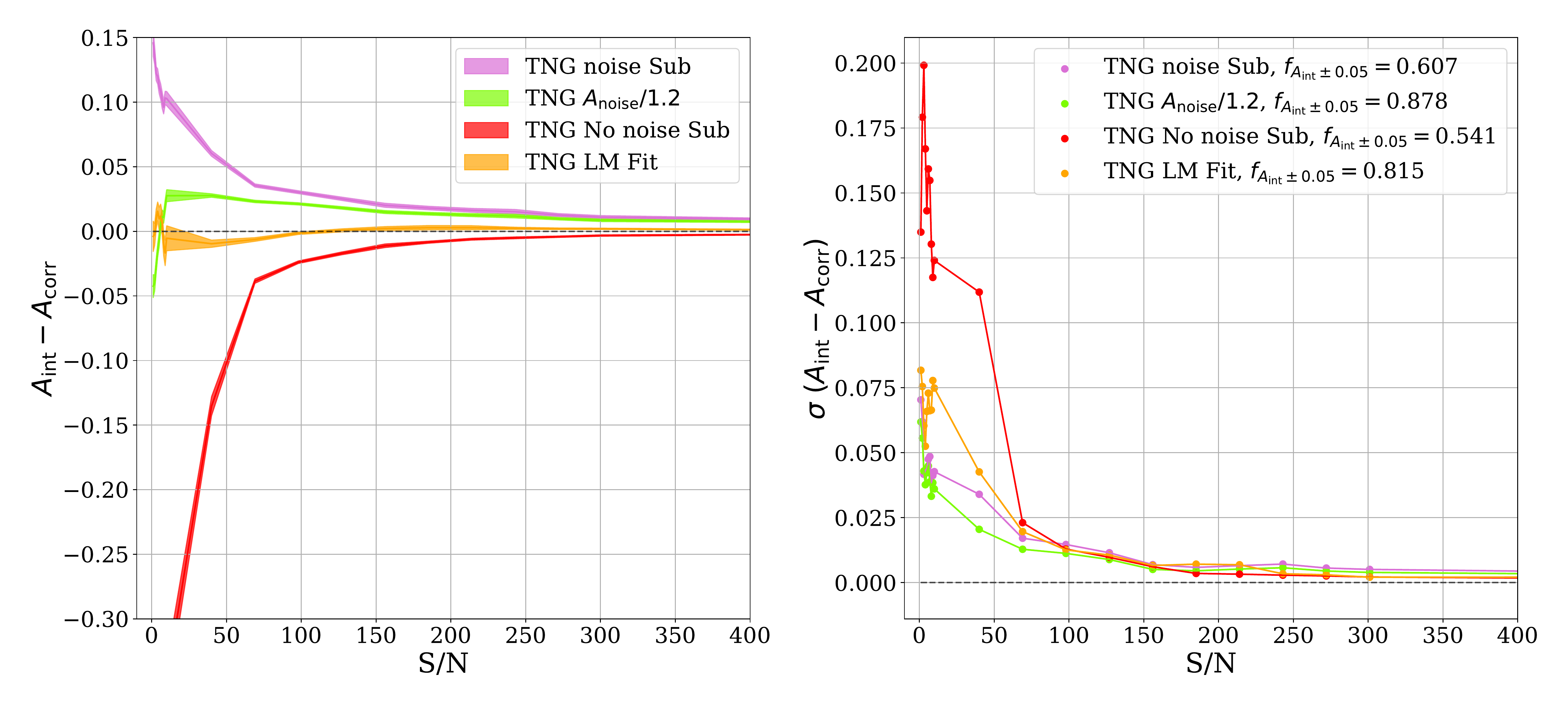}
	\centering
    \caption{Left: The mean difference between the intrinsic and observed asymmetry for increasing signal-to-noise bins. The red line represents performing no correction for the sky asymmetry, the purple line represents correcting for $A_{\rm noise}$, the green line for correcting $A_{\rm noise}/1.2$, and the orange line for applying the fit correction from Equation \ref{eqn:A_noise_fit}. The width of each line is the standard error on the mean. Right: The standard deviation of the difference in asymmetry for the four asymmetry correction methods. All methods that perform some kind of correction for noise contribution to asymmetry result in lower standard deviations than doing no noise subtraction. One should note that the standard deviation of the difference for the LM Fit method is comparable to the two noise subtraction methods except at low signal to noise. Here the scatter of the LM Fit method is slightly larger, though still relatively small and a significant improvement from doing no noise correction.}
    \label{fig:TNG_A_corr}
\end{figure*}

To determine the best method for measuring asymmetry we can compare the results from Equation \ref{eqn:A_noise_fit} to the traditional noise subtraction, the noise subtraction divided by 1.2, and doing no correction for noise asymmetry. Figure \ref{fig:TNG_A_corr} examines how well each of these methods reproduce the intrinsic asymmetry in different signal-to-noise regimes. The traditional method of subtracting the noise asymmetry (the purple line) does poorly at low signal-to-noise, as expected based on the earlier discussions in this work. By excluding galaxies with $S/N<100$, we can guarantee this bias in asymmetry is no more than 0.02, which given the range of possible asymmetry values (0-1) is relatively small. Even for low asymmetry galaxies where the bias has a greater percentage effect, because the bias is relatively constant after $S/N=100$ you could still compare asymmetries within a sample of galaxies with this $S/N$ cut. The signal-to-noise cut must be made though, otherwise the asymmetry of faint objects will appear systematically higher than brighter, nearby objects.

However, few galaxy properties (like those collected from an IFU observations) can achieve high enough signal-to-noise to recovery the majority of intrinsic asymmetry values with all of these correction methods. Though the signal-to-noise of the flux emission lines measured from spectra are excellent by design, other uncertainties that go into computing properties like stellar mass surface density can raise the statistical uncertainty of the pixel up to 0.1 dex or more. If one of our TNG stellar mass maps had an error of 0.1 dex on every pixel, the signal-to-noise of that stellar mass map (based on our calculation method) would be approximately 4. Thus, if we want to examine the asymmetry of spectral data products, the $A_{\rm noise}/1.2$ method is the better traditional noise correction compared to the $A_{\rm noisesub}$ method. The bias reaches less than 0.025 difference for $S/N$ as low as 50, and no bias exists at $S/N$>100. The asymmetry fit method, on the other hand, does better than all other asymmetry approximations at low $S/N$, with a difference between $A_{\rm int}$ and $A_{\rm obs}$ below 0.02 even at signal-to-noise$<$10, making it the ideal asymmetry measurement for data products with large uncertainties. This method has the added benefit of also not requiring a measurement of $A_{noise}$, so long as a global $S/N$ value can be approximated accurately. The relationship between observed asymmetry, signal-to-noise, and intrinsic asymmetry is fit to this simulation for a particular subset of galaxies; in Section \ref{sec:compare} we evaluate how well this fit can be applied to other datasets by comparing to Illustris.

\subsection{Spatial Resolution}
\label{sec:psf}

\begin{figure*}
	\includegraphics[width=0.9\textwidth]{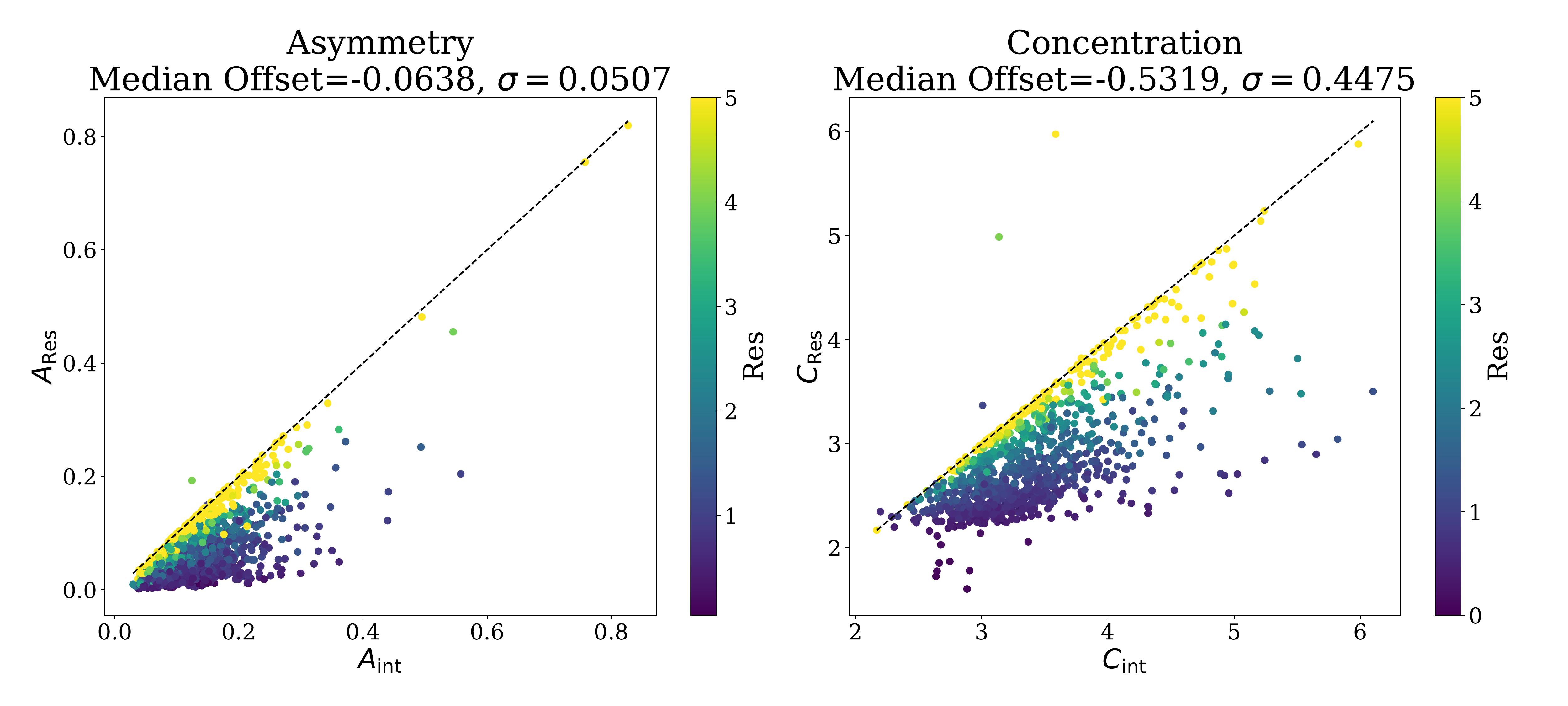}
	\centering
    \caption{A demonstration of how degrading resolution changes the asymmetry (left) and the concentration (right) compared to their intrinsic values (where resolution is functionally infinite). Degrading the resolution systematically decreases both asymmetry and concentration, and the magnitude of this difference increases inversely with the value of resolution. Note that the points with the lowest resolution are furthest from the black dashed line representing $A_{\rm Res}$=$A_{\rm int}$. Given the bias in the asymmetry/concentration measurement is clearly dependent on the intrinsic asymmetry/concentration and resolution value, just as with $S/N$ we should be able to correct for resolution effects and recover the intrinsic index.}
    \label{fig:PSF_change}
\end{figure*}

The resolution of an image can alter both the asymmetry and concentration measurements (e.g. Figure \ref{fig:Ex_Noise_PSF}). Low resolution can smooth out asymmetric structures that would otherwise lead to a larger asymmetry measurement. Degrading the resolution of an image will also spread out the light distribution (or stellar mass distribution in this case), resulting in more light in the outer aperture compared to the inner aperture than in the original image. Thus, degrading the resolution will lead to a systematically lower concentration as well. Figure \ref{fig:PSF_change} shows both of these effects for our sample of TNG galaxies. In the left panel the asymmetry is systematically under-estimated as the resolution decreases, and the right panel shows a similar under-estimation of concentration with decreasing resolution. The effect of varying resolution when comparing asymmetry and concentration values has been discussed in detail in previous works \citep{Bershady2000StructuralSample, Conselice2000TheGalaxies,Lotz2004AClassification,Bendo2007VariationsSequence,Giese2016Non-parametricLopsidedness}. 

\subsubsection{Resolution Corrected Asymmetry}
\label{sec:Res_A}

Just as we corrected the asymmetry for noise based on a relationship between $S/N$ and $A_{\rm obs}$ (Equation \ref{Eqn:A_int_blank}), we can assume a similar relationship exists between $A_{\rm Res}$ and the resolution of the image that can recover $A_{\rm int}$. Though we define our resolution as the stellar half mass radius over the PSF FWHM (both in arcseconds), any approximation of the apparent galaxy size in respect to the PSF size would suffice here. Using the Res parameter instead of the PSF FWHM alone guarantees the correction will work for a variety of galaxy sizes and distances, rather than just working for a range of observational PSFs. Given that resolution and $S/N$ have a similar effect on asymmetry, we assume a similar relationship between $A_{\rm int}$, $A_{\rm Res}$, and Res exists of the following form:

\begin{equation}
    \begin{aligned}
        & A_{\rm int} = (F_1(\rm Res)+1)A_{\rm Res}+F_2(\rm Res)A_{\rm Res}^2 +F_3(\rm Res)A_{\rm Res}^3  \\
    \end{aligned}
    \label{eqn:A_fit_PSF_blank}
\end{equation}

\noindent where $A_{\rm Res}$ is the observed asymmetry with degraded resolution (note that $A_{\rm Res}$ has no noise contribution) and $F_3$, $F_2$, and $F_1$ are each unique functions of this resolution parameter "Res" of the following form:

\begin{equation}
    \begin{aligned}
    & F(\rm Res) = X_1(\rm Res)^{-1}+X_2(\rm Res)^{-2}+X_3(\rm Res)^{-3}
    \end{aligned}
    \label{Eqn:Polynomial_FWHM}
\end{equation}

\noindent As with Equation \ref{Eqn:A_int_blank}, we design the relationship such that $A_{\rm int}=A_{\rm Res}$ as Res goes to infinity. As was completed with signal-to-noise in Section \ref{sec:noise}, we fit $A_{\rm Res}$ and Res to Equation \ref{eqn:A_fit_PSF_blank} using LM-Fit, resulting in the following relationship:

\begin{equation}
    \begin{aligned}
        & A_{\rm int} = ((3.626\pm0.108)\rm Res^{-1}+(0.0289\pm0.0714)\rm Res^{-2} \\ & -(0.00680\pm0.00833)\rm Res^{-3} +1)A_{\mathrm{Res}} + (-(15.53\pm1.46)\rm Res^{-1} \\ & -(21.17\pm2.32)\rm Res^{-2}+(0.036\pm0.454)\rm Res^{-3})A_{\mathrm{Res}}^2 \\ & +((10.34\pm1.68)\rm Res^{-1}+(77.4\pm11.5)\rm Res^{-2} \\ & +(35.4\pm14.7)\rm Res^{-3})A_{\mathrm{Res}}^3 \\
    \end{aligned}
    \label{eqn:A_fit_PSF}
\end{equation}

\begin{figure}
	\includegraphics[width=0.9\columnwidth]{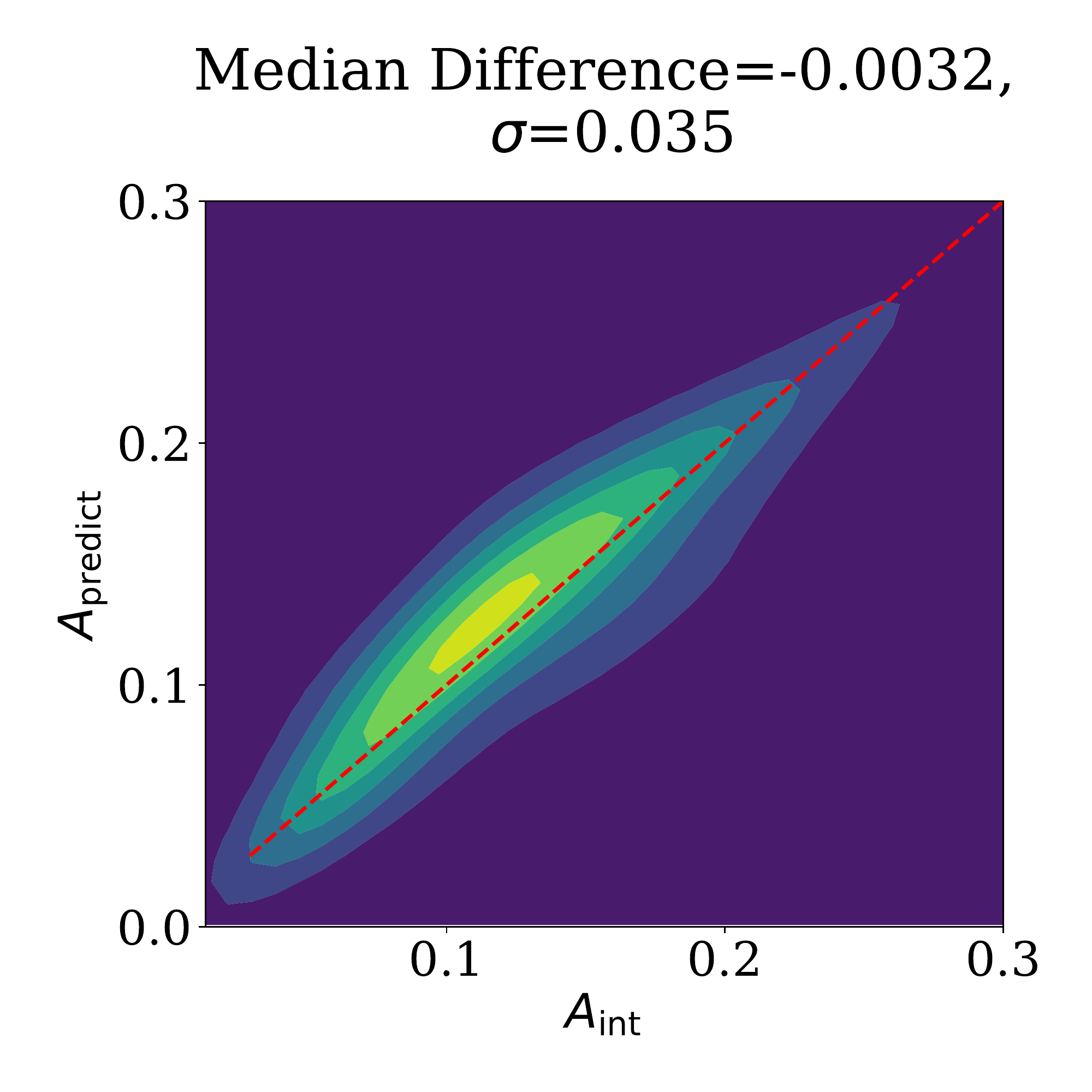}
	\centering
    \caption{Density contours of the asymmetry predicted by Equation \ref{eqn:A_fit_PSF}, plotted against the intrinsic asymmetry, with the red dashed line representing where the two would be equal. Note the median difference between these two values, even for the smallest asymmetries measurements, is only a few percent of the measured value. Thus the intrinsic asymmetry is recovered (within reason) for the majority of the galaxies in our sample.}
    \label{fig:A_PSF_corr_fit}
\end{figure}

Figure \ref{fig:A_PSF_corr_fit} confirms that the asymmetry predicted by Equation \ref{eqn:A_fit_PSF} ($A_{\mathrm{predict}}$) accurately replicates $A_{\rm int}$ for the majority of the sample, with a median difference that equates to only a few percent of the measured asymmetry (for even the lowest asymmetry galaxies). Equation \ref{eqn:A_fit_PSF} could be used to correct for very high $S/N$ observations, but in practicality resolution will need to be corrected for in conjunction with noise.

\begin{figure*}
	\includegraphics[width=0.9\textwidth]{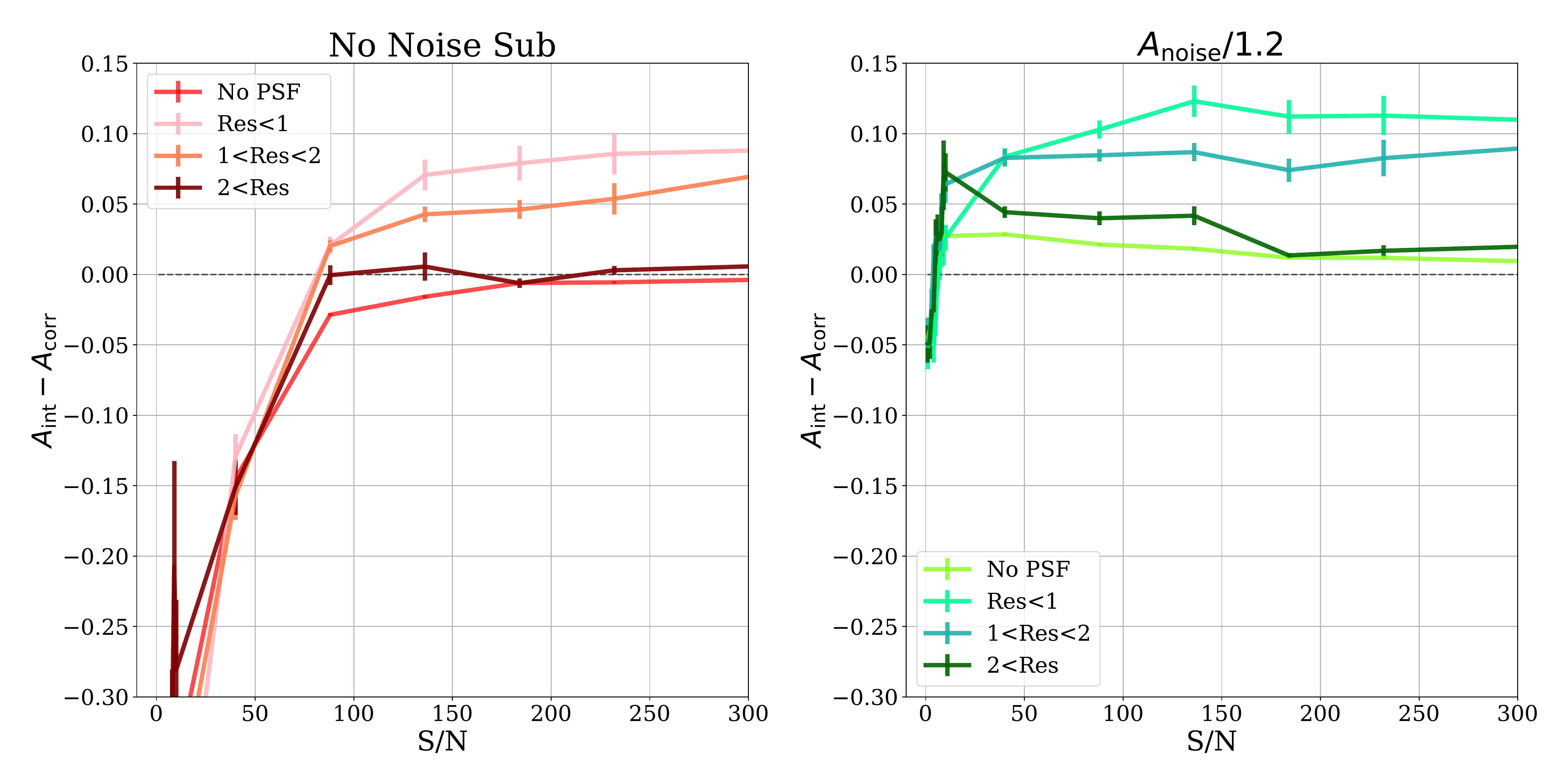}
	\centering
    \caption{A demonstration of how applying a PSF (and thus degrading the resolution) changes two of the asymmetry approximation methods from Figure \ref{fig:TNG_A_corr}. Left panel: How resolution changes the difference in asymmetry when no correction for the noise asymmetry is performed. The curve from Figure \ref{fig:TNG_A_corr} is replicated here, labelled "No PSF". The other three curves represent the relationship between $A_{\rm int}$-$A_{\rm corr}$ for different resolution bins. Applying a PSF to $A_{\rm obs}$ always shifts the curve up, since $A_{\rm obs}$ will systematically increase. This shift upward is greater for smaller resolutions. Right panel: The same investigation, but for the $A_{\rm noise}/1.2$ subtraction method. The curve from Figure \ref{fig:TNG_A_corr} is replicated here, labelled "No PSF", where as the other three curves are separated into resolution bins. The same systemic shift upward is seen as in both methods. Not only that, but beyond $S/N=200$ the change in asymmetry is almost entirely due to worsened resolution. The fact that both no noise subtraction and subtracting $A_{\rm noise}/2$ have similar changes from resolution beyond $S/N>200$ implies that above that $S/N$ we need only correct for resolution effects on asymmetry.}
    \label{fig:A_PSF_bins}
\end{figure*}

Figure \ref{fig:A_PSF_bins} showcases how different resolution values would change the relationship between $A_{\rm int}-A_{\rm corr}$ and $S/N$ for the no noise subtraction and $A_{\rm noise}/1.2$ subtraction correction methods. The original curve (with no PSF convolution) from Figure \ref{fig:TNG_A_corr} is replicated here for reference. Although the impact in asymmetry is seen most strongly at low signal-to-noise, at $S/N>100$ the change in asymmetry is dominated by the systematic decrease that comes from degrading resolution. The shift from decreasing resolution is also similar between the two correction methods, despite he $A_{\rm int}$-$A_{\rm corr}$ relationship differing between the two. The consistent upward shift of the curves above $S/N$=200 implies that there is some predictable change in asymmetry from PSF that is independent from signal-to-noise, and hence Equation \ref{eqn:A_fit_PSF} could be used to correct the asymmetry index irrespective of any choice of noise correction method. For $S/N<200$, the effects of noise and resolution are more entangled (almost indistinguishable for $S/N<100$) and need to be corrected for simultaneously.

To address the disparity between resolution changes in high and low $S/N$ regimes, we recommend a piecewise correction to asymmetry. For $S/N>200$ the asymmetry only needs to be corrected for resolution using Equation \ref{eqn:A_fit_PSF}, given the change in asymmetry from noise is comparatively so small. For $S/N<200$ we need to first correct for noise using Equation \ref{eqn:A_noise_fit}. The resulting $A_{\rm corr}$ still has effects from resolution, so we use LM-Fit determine the relationship between $A_{\rm corr}$, Res, and $A_ {\rm int}$ assuming they follow the same relationship as Equation \ref{eqn:A_fit_PSF_blank}. If $A_{\rm corr}$ is the predicted asymmetry calculated from Equation \ref{eqn:A_noise_fit} using $A_{\rm Res + Noise}$ and $S/N$, the following equation describes a combined noise and resolution correction for asymmetry:

\begin{equation}
    \begin{aligned}
    &   \textbf{S/N<200: } A_{\rm int} = ((1.930\pm0.118)\mathrm{Res}^{-1} -(0.027\pm0.060)\mathrm{Res}^{-2} \\ & -(0.012\pm0.007)\mathrm{Res}^{-3}+1)A_{\rm corr} + ((-15.209(\pm1.260)\mathrm{Res}^{-1} \\ & -(3.000\pm0.922)\mathrm{Res}^{-2} + (0.307\pm0.127)\mathrm{Res}^{-3})A_{\rm corr}^2  \\ & + ((23.629\pm3.220)\mathrm{Res}^{-1} +(11.098\pm3.000)\mathrm{Res}^{-2} \\ & - (1.042\pm0.454)\mathrm{Res}^{-3})A_{\rm corr}^3\\
    &   \textbf{S/N>200: } A_{\rm int} = \mathrm{Equation } \ref{eqn:A_fit_PSF}
    \end{aligned}
    \label{eqn:A_piecewise}
\end{equation}

\begin{figure}
	\includegraphics[width=0.87\columnwidth]{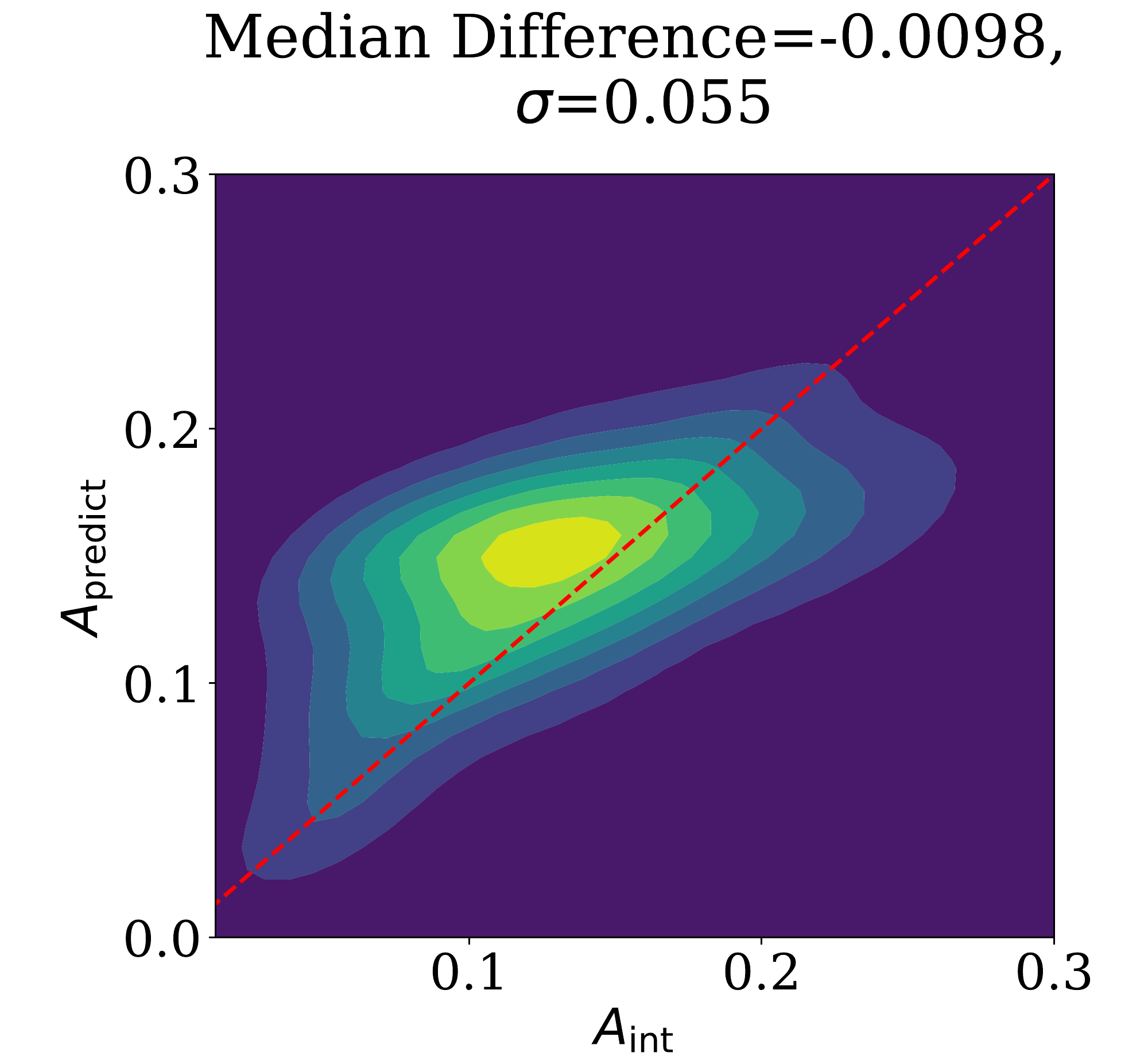}
	\centering
    \caption{Density contours of the asymmetry predicted by Equation \ref{eqn:A_piecewise} to correct for resolution and noise, plotted against the intrinsic asymmetry. The red dashed line represents where the two values would be equal to each other. The median difference is only a small fraction of the intrinsic asymmetry, though it is larger that the corrections for noise (Figure \ref{fig:Fit_scat}) and resolution (Figure \ref{fig:A_PSF_corr_fit}) alone. Correcting for noise and resolution simultaneously is inherently more difficult because both artefacts change asymmetry in opposite directions, making it difficult to distinguish which fraction of the change in asymmetry stems from noise or resolution. Most of the galaxies that constitute the outer contours have both a small signal-to-noise and resolution, demonstrating a weakness of the combined fit correction for the worst quality asymmetries. However, a median offset of 0.0098 is still good enough quality to make decent asymmetry measurements for the majority of the sample.}
    \label{fig:A_piecewise_correct}
\end{figure}

Figure \ref{fig:A_piecewise_correct} demonstrates how well Equation \ref{eqn:A_piecewise} recovers the intrinsic asymmetry. Comparing to the isolated noise (see Figure \ref{fig:Fit_scat}) and resolution (see Figure \ref{fig:A_PSF_corr_fit}) corrections, the median offset when correcting for both simultaneously is larger by a factor of 10, with an increased scatter as well. A slightly worse fit is to be expected, considering resolution and noise change asymmetry in opposing ways. Given the effects of noise and resolution on asymmetry can cancel each other out, it makes it very difficult for a fit based on $A_{\mathrm{Res}+\mathrm{noise}}$ to determine the individual change from resolution or noise alone. The overall broadening of the distribution of $A_{\mathrm{predict}}$ vs $A_{\mathrm{int}}$ stems from the dependence of the decrease in asymmetry from resolution effects on the intrinsic asymmetry: the absolute change in asymmetry from degrading resolution cannot be greater than its initial asymmetry value. Thus when the fit correction does poorly on small $A_{\mathrm{int}}$, resolution has a smaller effect than the noise contribution, which will drive an overestimation in $A_{\mathrm{predict}}$. When the fit does poorly on a large $A_{\mathrm{int}}$, the effects of resolution will dominate and lead to an underestimated $A_{\mathrm{predict}}$. Be that as it may, the broadening is driven entirely by galaxies with a combined low $S/N$ and Res value (roughly $S/N<20$ and Res$<1$). For better measurements the combined fit does well enough that on average the combined correction is sufficient for the work herein. Future analysis could be devoted to improving the joint correction, using other advanced fitting techniques such as machine learning algorithms to disentangle the two effects for even the smallest $S/N$ and Res values.

\begin{figure*}
	\includegraphics[width=0.9\textwidth]{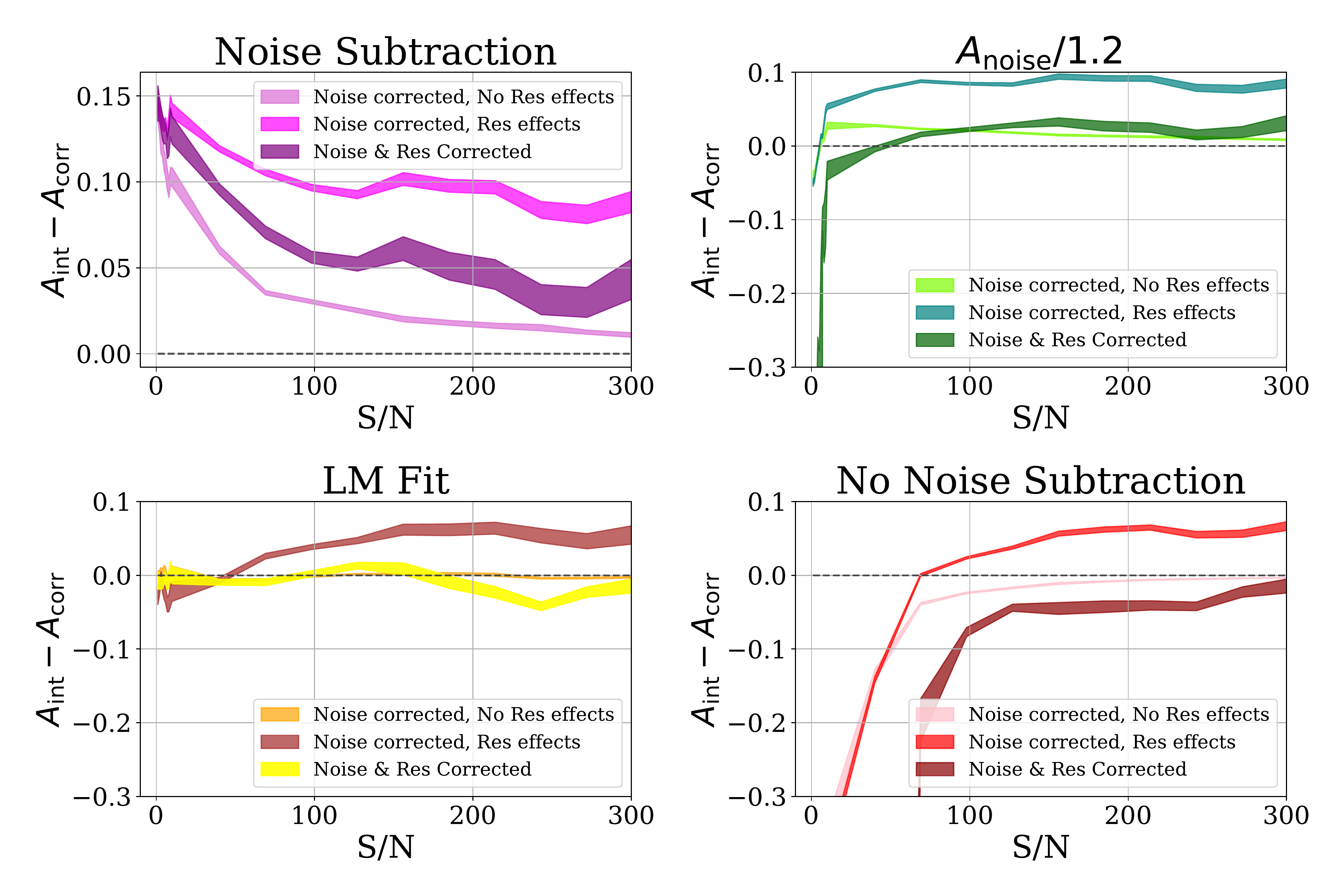}
	\centering
    \caption{The difference between intrinsic asymmetry and the asymmetry measured for four different noise correction methods with no resolution effects (the same as in Figure \ref{fig:TNG_A_corr}), with resolution effects, and after resolution correction. Given noise and resolution effect asymmetry in ways that can cancel out, it is difficult to disentangle the two observational effects. Thus the resolution corrected curve approximates the curve with no PSF, but not exactly (particularly at low $S/N$). The LM Fit method reproduces the intrinsic asymmetry at low $S/N$ most accurately from the four methods.}
    \label{fig:A_PSF_corr_methods}
\end{figure*}

We demonstrate how correcting for noise and PSF simultaneously with our LM-Fit equations is superior to correcting them separately in Figure \ref{fig:A_PSF_corr_methods}. Each panel represents a different noise correction method as presented in Section \ref{sec:noise}, with the original curves from Figure \ref{fig:TNG_A_corr}. When we degrade the resolution of the image and try to do the same correction (be it subtracting a fraction of the noise asymmetry or predicting asymmetry from Equation \ref{eqn:A_noise_fit}), the curves are shifted upwards due to the systematically lower asymmetry measurement once the resolution is degraded. For the "noise subtraction", "$A_{\rm noise}/1.2$", and "no noise subtraction", use Equation \ref{eqn:A_fit_PSF} to predict the intrinsic asymmetry form the noise corrected asymmetry and the Res value. Doing this brings us closer to the original curve, but there is still some disparity likely stemming from Equation \ref{eqn:A_fit_PSF} being constructed from asymmetries with no noise contribution and the slightly imperfect nature of these three noise correction methods. The curves are better recovered above $S/N=100$ than below, supporting the conclusion that noise must be correctly accounted for to use Equation \ref{eqn:A_fit_PSF} on more realistic images. 

The "LM Fit" method, on the other hand, uses Equation \ref{eqn:A_piecewise} to correct for resolution and noise simultaneously. In this case, the Noise+PSF corrected curve almost completely overlaps the original noise correction curve with no resolution effects. Though Figure \ref{fig:A_PSF_corr_methods} demonstrates that a resolution correction fit like Equation \ref{eqn:A_fit_PSF} can be applied to noise correction methods that do not require a fit, the most accurate way to account for both noise and resolution is to correct them with a fit function simultaneously (Equation \ref{eqn:A_piecewise}).

\subsubsection{Resolution Corrected Concentration}

\begin{figure}
	\includegraphics[width=0.8\columnwidth]{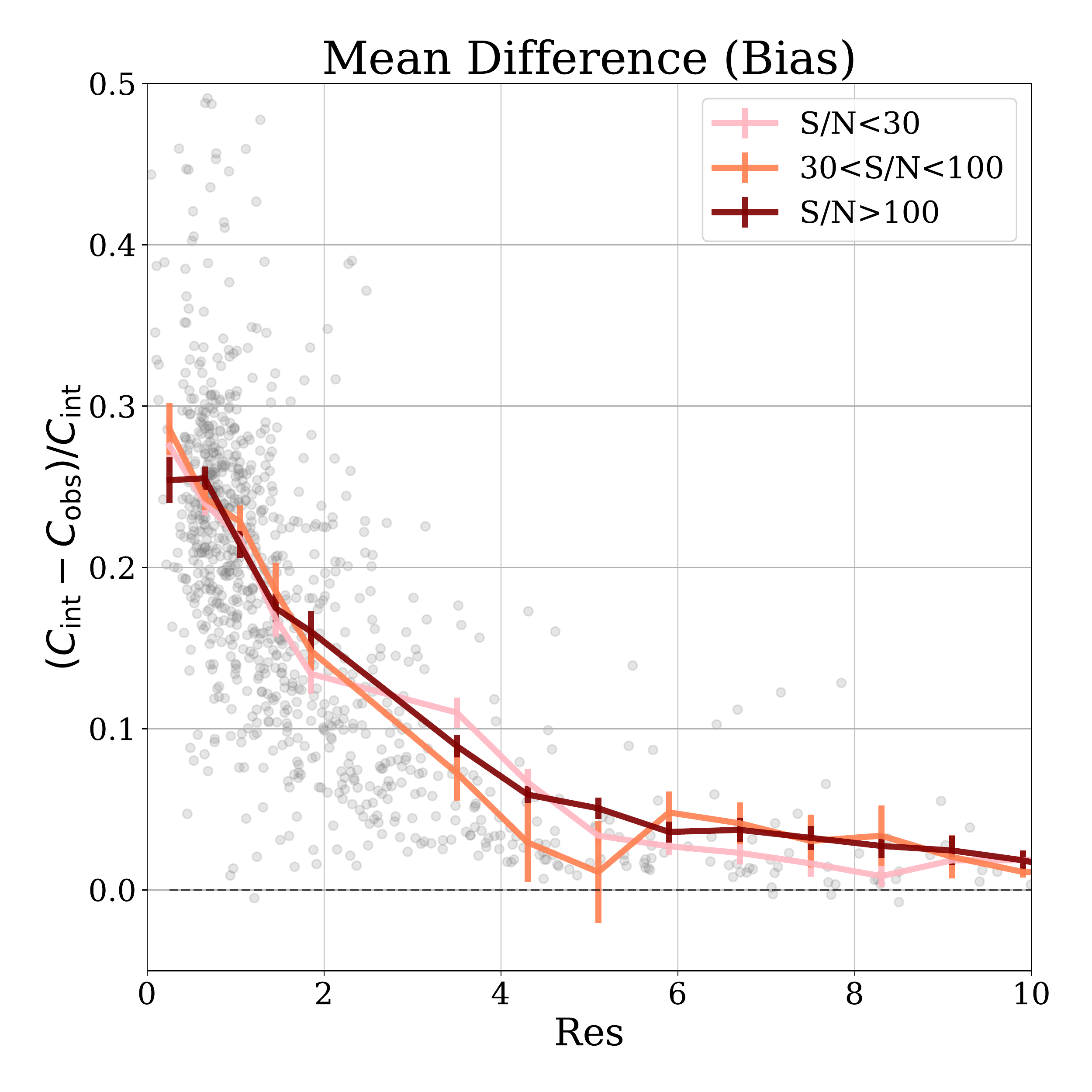}
	\centering
    \caption{The mean fractional difference between intrinsic and observed concentration as a function of resolution for different $S/N$ bins. The three curves overlap each other, indicating that a single resolution correction can be applied to concentrations of varying $S/N$. This logically follows the results from Figure \ref{fig:Int_Noise}, which demonstrated that noise has a negligible effect on the concentration measurement.}
    \label{fig:A_PSF_corr_TNG_app}
\end{figure}

In Section \ref{sec:noise} we demonstrated that concentration is unaffected by noise. Figure \ref{fig:A_PSF_corr_TNG_app} shows that the relationship between the fractional change in concentration as a function of resolution is unchanged for different signal-to-noise bins. I.e., a resolution value will have the same effect on concentration for both a high and low signal-to-noise galaxy. This uniformity is expected considering how little adding random noise changes the concentration. Therefore a function can be fit to this data to correct any concentration underestimation from resolution, without also needing to account for noise correction (unlike asymmetry). We discuss the limitations of the LM-fit method further in Section \ref{sec:compare}.

Just as with the asymmetry correction for noise, we postulate that the relationship between $C_{\rm int}$, $C_{\rm Res}$, and Res=$R_{\rm half}$/FWHM can be described by a single function (equation \ref{eqn:C_fit_blank}).

\begin{equation}
    \begin{aligned}
    & C_{\rm int} = (F_1(\rm Res)+1)C_{\rm Res}  + F_2(\rm Res)C_{\rm Res}^2 + F_3(\rm Res)C_{\rm Res}^3 
    \end{aligned}
    \label{eqn:C_fit_blank}
\end{equation}

\noindent Where $F_3$, $F_2$, and $F_1$ are of the same form as Equation \ref{Eqn:Polynomial_FWHM}. Once again we have constructed the relationship to ensure that $C_{\rm int}=C_{\rm Res}$ as the resolution goes to infinity. Utilising LM-Fit, we find that the best equation to express this relationship is: 

\begin{equation}
    \begin{aligned}
    & C_{\rm int} = (-(3.616\pm0.296)\rm Res^{-1} - (1.467\pm0.268)\rm Res^{-2} \\ & -(0.6106\pm0.0716)\rm Res^{-3} + 1)C_{\rm Res} + ((2.159\pm0.164)\rm Res^{-1} \\ & + (0.996\pm0.225)\rm Res^{-2} + (0.7367\pm0.0848)\rm Res^{-3})C_{\rm Res}^2 \\ & + (- (0.3090\pm0.0226)\rm Res^{-1} - (0.0825\pm0.0432)\rm Res^{-2} \\ & -(0.2197\pm0.0247)\rm Res^{-3})C_{\rm Res}^3  \\
    \end{aligned}
    \label{eqn:C_fit}
\end{equation}

\begin{figure}
	\includegraphics[width=0.9\columnwidth]{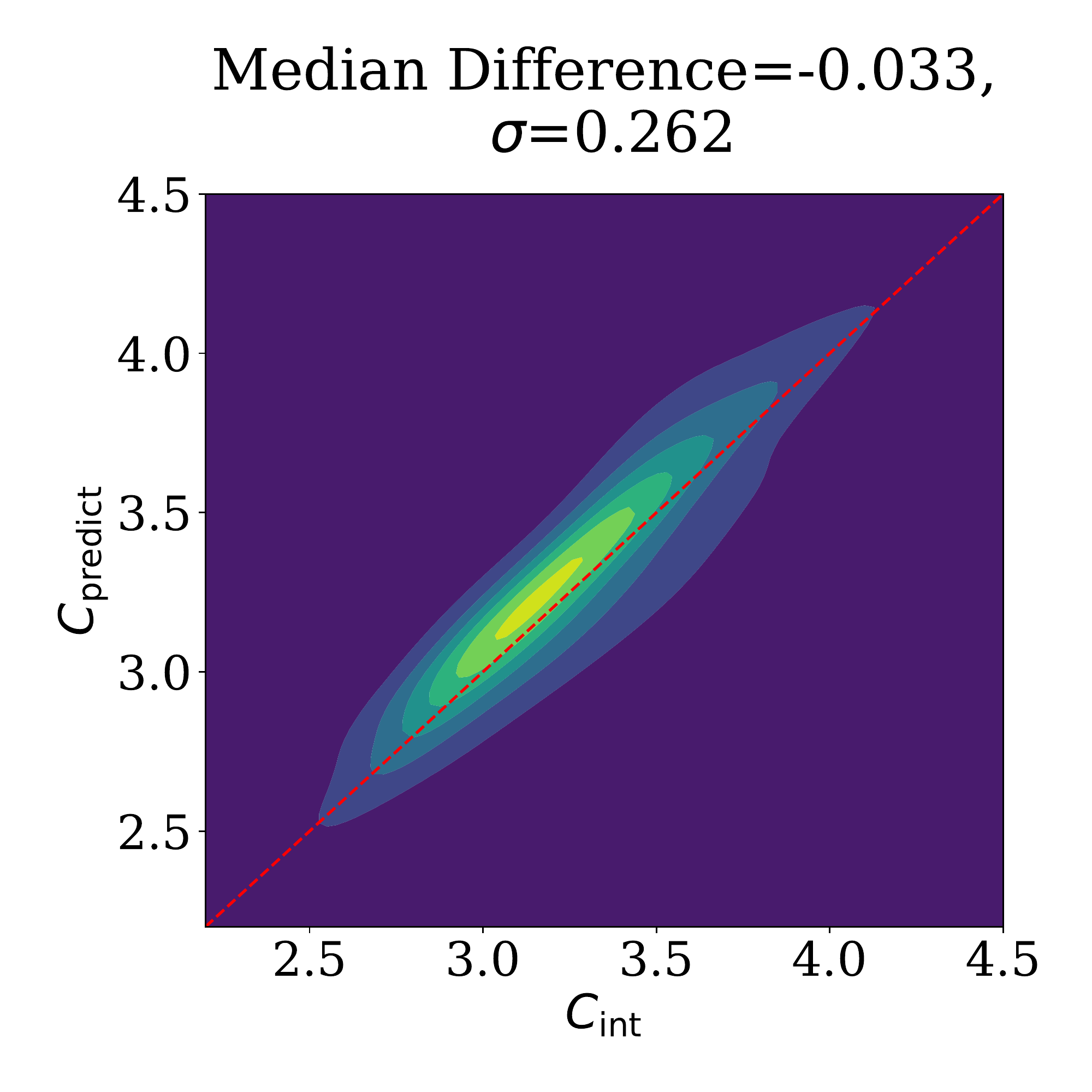}
	\centering
    \caption{Density contours for the concentration predicted by Equation \ref{eqn:C_fit} versus the intrinsic concentration, with the red dashed line representing $C_{\rm predict}$=$C_{\rm int}$ The small, unbiased scatter about the line of equity demonstrates how effective the resolution correction is for this simulation. The median difference in concentration is less than 1$\%$ of the average measured concentration.}
    \label{fig:C_PSF_fit}
\end{figure}

\noindent This function predicts concentration values that are, on average, within 1$\%$ of the intrinsic concentration value (with a median difference of -0.033). Figure \ref{fig:C_PSF_fit} further demonstrates that the predicted and intrinsic concentration are tightly correlated.

Thus, both asymmetry and concentration can be reliably corrected for resolution as low as 0.05. Such a correction is crucial to allow comparison of non-parametric morphology measurements for different resolutions, particularly if one wants to compare the morphologies of different observational data sets. The accuracy and convenience of a resolution correction is crucial in particular for asymmetry, given the more complex choices required to correct asymmetry for noise. In the next section we evaluate the general applicability of the resolution and noise correction we have outlined thus far, by testing them on unseen TNG data and the Illustris simulation.

\section{Discussion}
\label{sec:discussion}

\subsection{Illustris Comparison}
\label{sec:compare}

In Section \ref{sec:analysis}, we demonstrated how by fitting the relationship between observed and measured asymmetry/concentration, it is possible to correct for most of the bias created by an observational effect such as noise/resolution. For asymmetry, this fitting method performs better than other correction methods we investigate for background noise. In particular, the fitting method allows for correction of asymmetry measurements down to very low $S/N$. However, before such a fitting method can be used in practice, we must check its universality, or whether its form is appropriate only for the sample of TNG galaxies used thus far.

We first check that Equations \ref{eqn:A_noise_fit}, \ref{eqn:A_fit_PSF}, \ref{eqn:A_piecewise}, and \ref{eqn:C_fit} are not overfit to the training sample by applying those corrections to a set of test TNG galaxies not included in our initial 1000 galaxy set from which the fits were determined. We select 1000 new stellar mass maps from TNG galaxies, using the same quality cuts described in Section \ref{sec:TNG}, but only selecting galaxies we have not used in the analysis yet. We select the 1000 galaxies randomly, and thus the global properties (such as total stellar mass and stellar half mass radius) of the test set are comparable to our initial TNG sample. We then add noise from the same range of input $S/N$ values, and convolve with a PSF to achieve the same range of resolution values (refer back to Section \ref{sec:CAS} for details). We can test to see how well Equations \ref{eqn:A_noise_fit}, \ref{eqn:A_fit_PSF}, \ref{eqn:A_piecewise}, and \ref{eqn:C_fit} recover the intrinsic asymmetry/concentration values for this new dataset. 

\begin{figure*}
    \centering
    \subfloat[Unseen TNG Galaxies]{\includegraphics[width=0.9\textwidth]{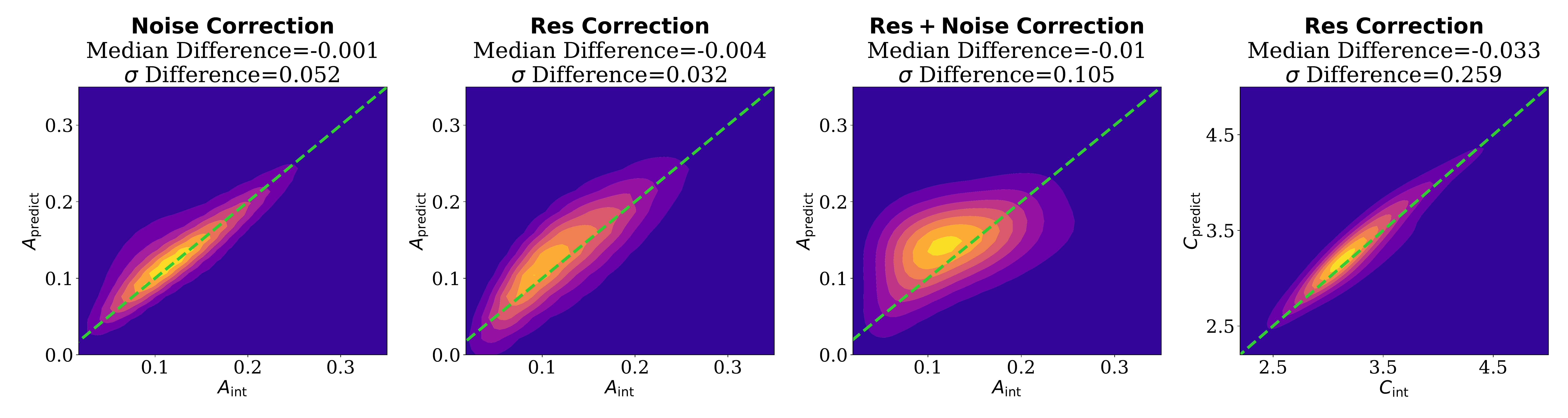}}%
    \qquad
    \subfloat[Illustris Galaxies]{\includegraphics[width=0.9\textwidth]{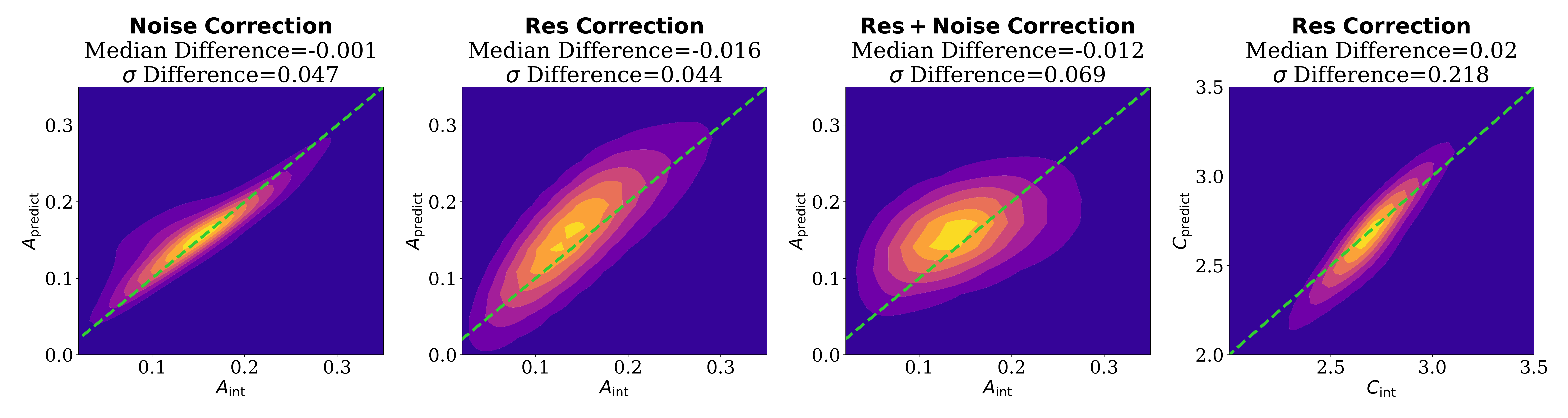}}%
    \caption{Top row: Density contours of the intrinsic morphologies compared to values predicted by the fits presented in this work for a set of previously unseen TNG galaxies, with the dashed green line representing where the predicted values equal the intrinsic value. The first panel shows the asymmetry predicted after noise correction, the 2nd shows asymmetry predicted after resolution correction, the 3rd shows asymmetry corrected for both resolution and noise, and the fourth shows concentration predicted after resolution correction (no noise correction for concentration is necessary). Bottom row: Density contours of the intrinsic morphologies compared to values predicted by the fits presented in this work for a set of Illustris Galaxies, with the dashed green line representing where the predicted values equal the intrinsic value. The order of whether noise, resolution, or a combination of both is corrected for is the same as in the top row. For the combined noise and resolution correction (third column) galaxies with asymmetries predicted to be greater than 1 (far beyond the scope of $A_{\rm int}$) are removed from the analysis (20 in the unseen TNG and 4 in Illustris). For both datasets there is no clear bias in the predicted values, and the scatter on the predicted values reflects that seen in the training data (see Figures \ref{fig:Fit_scat}, \ref{fig:A_PSF_corr_fit}, \ref{fig:A_piecewise_correct} and \ref{fig:C_PSF_fit}). For comparison the initial median offset for noise correction of asymmetry is -0.001, for resolution correction of asymmetry is -0.0032, for resolution and noise corrected asymmetry is -0.0098, and for resolution correction of concentration is -0.033. The combined noise and resolution correction does have a broader distribution than the other corrections, where galaxies with both a small $S/N$ and Res value result in poorer asymmetry predictions. But the same effect is scene in the original TNG sample (see Figure \ref{fig:A_piecewise_correct}). The replicability of the difference between the intrinsic and predicted indices with both Illustris and TNG implies that our corrections are viable for different galaxy samples.}
    \label{fig:Check_overfit_TNG}
\end{figure*}

The top panel of Figure \ref{fig:Check_overfit_TNG} presents the quality of the predicted asymmetry for these new galaxies: correcting for noise using Equation \ref{eqn:A_noise_fit} (1st panel), for resolution using Equation \ref{eqn:A_fit_PSF} (2nd panel), and for resolution and noise using equation \ref{eqn:A_piecewise} (3rd panel). There are 20 galaxies (2\% of the sample) for which the asymmetry is predicted to be unrealistically high when correcting for noise and resolution together ($A_{\rm predict}>1$); we remove these from the rest of the analysis so the statistics measured reflect the behaviour of the majority of the sample rather than a few outliers. The 4th panel shows how well Equation \ref{eqn:C_fit} corrects the concentration for resolution. The difference between the predicted and intrinsic values for this new sample are congruent with those seen in the training sample (compare to Figure \ref{fig:Fit_scat} (median offset -0.001), Figure \ref{fig:A_PSF_corr_fit} (median offset -0.0032), Figure \ref{fig:A_piecewise_correct} (median offset -0.0098), and Figure \ref{fig:C_PSF_fit} (median offset -0.033) respectively). Thus there is no evidence that our equations are overfit to our initial galaxy sample. With that confirmed, we can now investigate the robustness of the fit: i.e., are Equations \ref{eqn:A_noise_fit}, \ref{eqn:A_fit_PSF}, \ref{eqn:A_piecewise}, and \ref{eqn:C_fit} robust enough to be applied to other simulations, or ultimately, to observational data?

Assessing the accuracy of our fitting method is not possible on observational data, given we cannot know the "intrinsic" asymmetry or concentration of an observed galaxy. However, we can use a different simulation suite as a test case, to see how well our suggested fitting corrections work on an entirely different set of data. For this, we choose to test our fit on a sample of 1000 galaxies from the Illustris-1 simulation \citep{2014Natur.509..177V,Genel2014IntroducingTime,Sijacki2015TheTime}. Though simulated using the same code as Illustris, TNG has an updated galaxy formation model which is known to affect the morphologies of the galaxies \citep{Rodriguez-Gomez2017TheMorphology,Rodriguez-Gomez2019TheObservations}. If the underlying physics in a simulation changes either the asymmetry and concentration correction fits, then it is unlikely we could apply those same fits to other simulations or observations. More importantly, testing on a different dataset will help further verify the signal-to-noise regimes in which the asymmetry fit correction is dependable.

\begin{figure*}
	\includegraphics[width=0.9\textwidth]{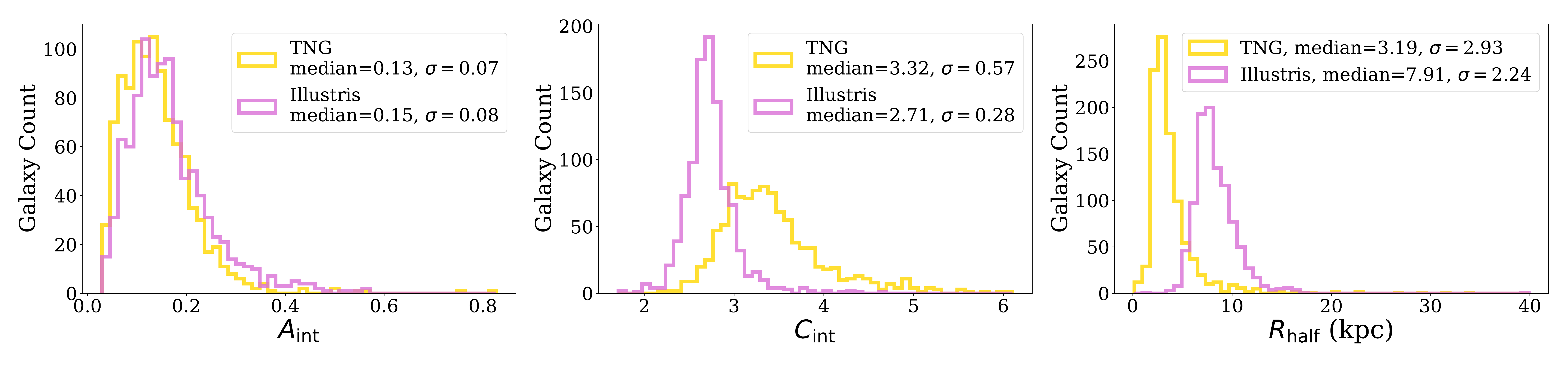}
	\centering
    \caption{Distributions of intrinsic asymmetry (left), intrinsic concentration (middle) and stellar half-mass radius (right) for the 1000 galaxies selected from TNG (yellow) and Illustris (pink) galaxies. Note that Illustris galaxies are on average larger and more highly concentrated, but have similar asymmetries as those in TNG. TNG has more accurate sizes for low mass galaxies (reflected in the smaller median $R_{\mathrm{half}}$ in TNG compared to Illustris), which likely leads to lower concentration values than were observed in Illustris.}
    \label{fig:AC_TNG_Il}
\end{figure*}

The middle and right panels of Figure \ref{fig:AC_TNG_Il} shows the differing distributions of asymmetry and concentration for the TNG and Illustris sub-samples used in this work, along with the stellar half-mass radius of those galaxies. Though the asymmetry distributions are relatively similar, TNG has more highly concentrated galaxies than Illustris. \cite{Rodriguez-Gomez2019TheObservations} noted this difference in concentration between TNG and Illustris, attributing it to the more accurate size of low-mass galaxies (see left panel of Figure \ref{fig:AC_TNG_Il}) caused by the treatment of galactic winds and the enhanced quenching of high-mass galaxies from the new AGN and stellar feedback model. Given that our correction methods require input from both the observed concentration/asymmetry and the applied resolution/noise, the difference in distributions should not alter the viability of a comparison between the two samples.

\begin{figure*}
	\includegraphics[width=0.9\textwidth]{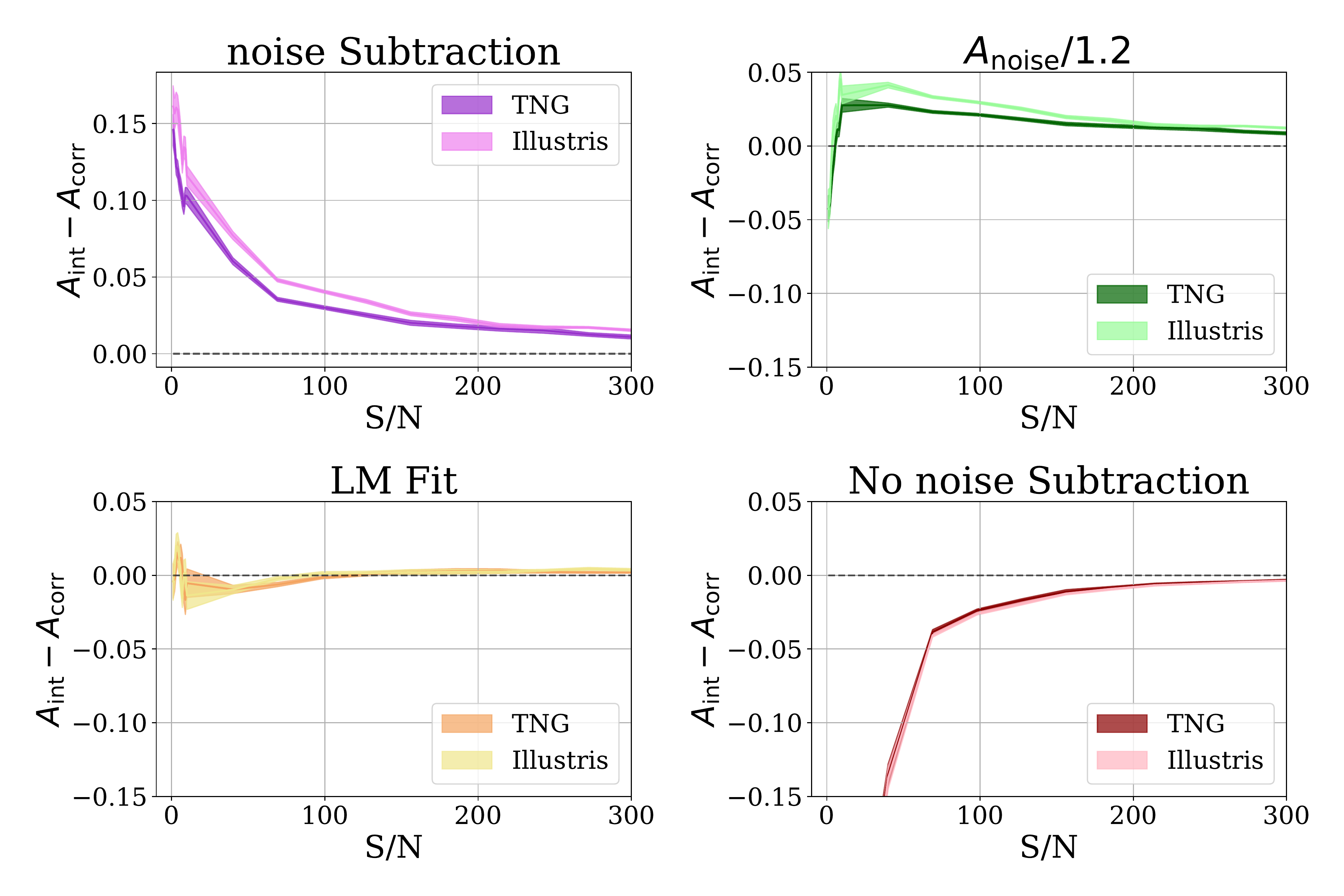}
	\centering
    \caption{The mean difference between the intrinsic and corrected asymmetry as a function of signal-to-noise for subtracting $A_{\rm noise}$ (top left), subtracting $A_{\rm noise}/1.2$ (top right), predicting asymmetry from Equation \ref{eqn:A_noise_fit} (bottom left), and doing no noise correction (bottom right) for both our TNG (the dark curve) and Illustris (the light curve) galaxy sets. All four correction methods match between the galaxy samples, implying that the relationship between the quality of correction and $S/N$ is the same no matter which simulation is examined.}
    \label{fig:TNG_Ill_A_corr}
\end{figure*}

We then apply the same noise addition (and range of $S/N$ values) to Illustris as described in Section \ref{sec:methods}, such that we can replicate the curves displayed in the left panel of Figure \ref{fig:TNG_A_corr} with Illustris, and compare them to the same results acquired with TNG (see Figure \ref{fig:TNG_Ill_A_corr}). Doing no noise correction, subtracting $A_{\rm noise}$, and subtracting $A_{\rm noise}/1.2$ are shown to all work equally well for both Illustris and TNG in all signal-to-noise regimes. Not only that, but because the relationship between $A_{\rm int}-A_{\rm corr}$ and signal-to-noise is almost identical between TNG and Illustris for these three methods, we could use those relationships to predict the offset in asymmetry for any correction method used with data of a particular signal-to-noise. The bottom left hand panel of Figure \ref{fig:TNG_Ill_A_corr} demonstrates how using Equation \ref{eqn:A_noise_fit} to correct Illustris asymmetries for noise works just as well it did on TNG, despite the data being fit to minimize $A_{\rm int}-A_{\rm corr}$ for TNG alone. The bottom left panel of Figure \ref{fig:Check_overfit_TNG} shows this more explicitly with the direct comparison of $A_{\rm predict}$ and $A_{\rm int}$, confirming that the median difference and scatter in the difference between the asymmetry predicted by Equation \ref{eqn:A_noise_fit} and the intrinsic Illustris asymmetry is comparable to that seen in TNG (see Figure \ref{fig:Fit_scat}).

Next we test the effects of resolution on Illustris, and how well Equations \ref{eqn:A_fit_PSF}, \ref{eqn:A_piecewise}, and \ref{eqn:C_fit} correct for them. Given the overall larger size of Illustris galaxies (see the distribution of $R_{\rm half}$ in the left panel of Figure \ref{fig:AC_TNG_Il}), we choose to select PSF FWHMs such that the Res values of Illustris match those in TNG (rather than applying the same range of PSFs to Illustris and TNG). If we applied PSFs ranging from 0.002"-7" FWHM to Illustris, they would have less effect on the overall resolution of Illustris galaxies than they did on TNG, given their greater size. In such a scenario we would only be testing the PSF correction for comparatively "good" resolutions. Instead we use the Res values in TNG to compute a new range of PSF values to apply to Illustris (ranging from 0.002"-30" FWHM), which results in the same effective resolution of galaxy structure in the two simulations (note that this is distinct from the resolution of the simulation itself).

We can use Equations \ref{eqn:A_fit_PSF} and \ref{eqn:C_fit} to correct for the changes to asymmetry/concentration on a Illustris galaxies with different resolutions. The bottom row of Figure \ref{fig:Check_overfit_TNG} demonstrates how well the predicted asymmetry (2nd panel) and concentration (4th panel) reflect the intrinsic values. The median difference and scatter in difference for both asymmetry and concentration is comparable to that for the predicted values in TNG (compare to Figures \ref{fig:A_PSF_corr_fit} and \ref{fig:C_PSF_fit}). The distribution of $C_{\rm predict}$ vs. $C_{\rm int}$ looks narrower than that in TNG, stemming from the narrower range and overall smaller $C_{\rm int}$ values in Illustris (see Figure \ref{fig:AC_TNG_Il}). We also demonstrate how a combined resolution and noise correction works on Illustris (Equation \ref{eqn:A_piecewise}) in the 3rd panel, where the median difference is similar to both that in the unseen TNG sample and the initial test sample (all are approximately off by 0.01, still well within a "good" asymmetry measurement).

The different physical models that shape TNG and Illustris galaxies do not alter the quality of the resolution and noise corrections we have described in this work. Thus, these corrections could feasibly be used on other datasets or even observations. The benefit of using simulations is that we implement any changes between the intrinsic and observed data, and thus understand it completely. The same care needs to be taken to understand the difference between an observational dataset and the simulation on which the fits are based. For example, we do not have to account for other observational effects that might add to (or detract from) the asymmetry outside of noise, such as neighbouring objects outside the field-of-view (when external flux leaks into the image of a galaxy). Our correction methods also depend on how $S/N$ and resolution are quantified; to use the fit coefficients provided in this work $S/N$ and Res need to be measured similarly to what is laid out in this analysis. Though the work herein stands as a proof of concept, we caution those wishing to use a fit correction to take into account the unique features of their galaxy sample, and consider creating a fit on simulations meant to emulate that sample.

\subsection{Implications}
\label{sec:implications}

There are a few studies that have attempted to measure non-parametric morphologies on data other than broad band light images, and how to account for the complex uncertainties in that scenario. \cite{Lanyon-Foster2012TheZ1} characterize the morphology of stellar mass maps determined from Hubble Space Telescope imaging using CAS as well as other morphological measurements. The asymmetry of their stellar mass maps was on average larger than from the optical light image, which they attribute in part to random fluctuations in the mass-to-light ratio. The smallest asymmetries they measure for stellar mass maps are negative, implying that the background asymmetry they estimate from both regular background fluctuations and a generalized contribution from mass-to-light ratio fluctuations is larger than the asymmetry of the galaxy. Both of these findings are in agreement with our results of high uncertainties leading to falsely large asymmetries, and subtracting the total $A_{\rm noise}$ resulting in an over-corrected asymmetry.

\cite{Giese2016Non-parametricLopsidedness} is the only other paper to suggest a fitting method to correct asymmetry for noise, whilst measuring the asymmetries of extended HI gas disks from the Westerbork observations of neutral Hydrogen in Irregular and SPiral galaxies (WHISP) sample \citep{2001ASPC..240..451V}. To account for the generally low $S/N$ measurements of HI, \cite{Giese2016Non-parametricLopsidedness} use the Tilted Ring Fitting Code (TIRIFIC) model to approximate a correction in asymmetry based on the $S/N$. Using machine learning they are able to approximate the intrinsic asymmetry within 0.05 for most galaxies. Rather than creating a universal correction method, they fine-tune their correction to the WHISP sample to further their study. Indeed, the problem with creating a fit based on a simulated galaxy sample is the need to calibrate simulated galaxies to reflect the observations of interest. This is particularly relevant if one is interested in measuring the asymmetry of a galaxy property like surface mass density, or star-formation rate density. In contrast, we have investigated the universality of multiple different noise correction methods, and just as importantly, the signal-to-noise regimes for which those methods are viable. Using the work herein one can evaluate an approximate accuracy of asymmetry and concentration measurements for a dataset and decide if any correction for bias in the measurements is necessary (and the most appropriate approach to correct for that bias).

\begin{table*}
\captionof{table}{Summary of the validity of different asymmetry noise correction methods in various $S/N$ regimes. For each method (no noise correction (``No corr''), subtract $A_{\rm noise}$ from the observed asymmetry (``Noise Sub''), subtract $A_{\rm noise}/1.2$ from the observed values (``$A_{\rm noise}/1.2$''), and corrected based on Equation \ref{eqn:A_noise_fit} (``LM Fit'')) we provide the mean difference and scatter in difference between the intrinsic asymmetry and the measured asymmetry. From this one can approximate the bias in asymmetry measurement for data with differing $S/N$ regimes. The signal to noise must be measured as laid out in this analysis, using the median within an aperture of 1$R_{\rm half}$ (an alternative radius with comparable size to $R_{\rm half}$ could be used as well). \label{tab:A_noise_corr_TNG}}
\begin{tabular*}{\textwidth}{c @{\extracolsep{\fill}} ccccccccc}
S/N  & No Corr Mean & No Corr $\sigma$ & Noise Sub Mean  & Noise Sub $\sigma$ & $A_{\rm noise}/1.2$ Mean & $A_{\rm noise}/1.2$ $\sigma$ & LM Fit Mean & LM Fit $\sigma$\\
\hline
S/N<5 & -0.772 & 0.167 & 0.122 & 0.044 & -0.027 & 0.043 & 0.009 & 0.066 \\
5<S/N<10 & -0.428 & 0.137 & 0.103 & 0.043 & 0.011 & 0.039 & 0.0 & 0.07 \\
10<S/N<25 & -0.349 & 0.124 & 0.103 & 0.043 & 0.028 & 0.036 & -0.005 & 0.075 \\
25<S/N<50 & -0.135 & 0.112 & 0.06 & 0.034 & 0.028 & 0.02 & -0.01 & 0.043 \\
50<S/N<75 & -0.039 & 0.023 & 0.035 & 0.017 & 0.023 & 0.013 & -0.006 & 0.02 \\
75<S/N<100 & -0.024 & 0.013 & 0.03 & 0.015 & 0.021 & 0.011 & -0.001 & 0.013 \\
100<S/N<150 & -0.017 & 0.01 & 0.025 & 0.011 & 0.018 & 0.009 & 0.001 & 0.011 \\
150<S/N<200 & -0.01 & 0.005 & 0.019 & 0.006 & 0.014 & 0.005 & 0.002 & 0.007 \\
200<S/N<250 & -0.006 & 0.003 & 0.016 & 0.007 & 0.012 & 0.005 & 0.003 & 0.005 \\
250<S/N<300 & -0.004 & 0.003 & 0.013 & 0.006 & 0.01 & 0.004 & 0.002 & 0.003 \\
300<S/N<400 & -0.003 & 0.002 & 0.011 & 0.005 & 0.009 & 0.004 & 0.002 & 0.002 \\
400<S/N<600 & -0.002 & 0.001 & 0.007 & 0.003 & 0.006 & 0.003 & 0.0 & 0.002 \\
600<S/N<900 & -0.001 & 0.001 & 0.005 & 0.002 & 0.004 & 0.002 & -0.001 & 0.001 \\
900<S/N<1200 & -0.0 & 0.0 & 0.004 & 0.002 & 0.003 & 0.002 & -0.0 & 0.0 \\
\end{tabular*}
\end{table*}

The analysis within this work provides the opportunity to approximate both the bias in asymmetry, and the scatter in that bias, created by different amounts of noise in an image such that future works can check the quality of asymmetry measurements for the dataset being used.
Table \ref{tab:A_noise_corr_TNG} provides a simplified version of the results from Figure \ref{fig:TNG_A_corr}, supplying both the mean offset and the standard deviation of that offset for the four noise-correction methods discussed in this work. Using Table \ref{tab:A_noise_corr_TNG}, one could check what the bias would be for a desired asymmetry correction method at the signal-to-noise of their particular set of observations, and determine which method would work best to both minimize the bias, as well as avoid varying biases within the same sample. As the community gets more creative with what kinds of observations and data products asymmetry is measured from, we hope this table can serve as a cautionary step as we explore this new regime of non-parametric morphologies.

It is paramount that when measuring asymmetry for a set of galaxies one can quantify the variability of an offset created by noise contribution. If one is working with high-quality photometry, or strong spectral lines, it is likely that the change in asymmetry from noise will be the same within that sample (no matter the noise-correction method chosen). However, comparisons of that data to lower-quality data sets becomes ill-advised, an issue which will only become more relevant as state-of-the-art non-parametric morphology measurements are compared to literature values. For low signal-to-noise data-sets it is even unwise to compare asymmetries within the data-set, given the bias created by a large noise/uncertainty can vary so drastically between asymmetry values. This is a particular issue for those measuring asymmetries in HI gas \citep[e.g][]{Holwerda2011QuantifiedGalaxies,Giese2016Non-parametricLopsidedness,Reynolds2020HGalaxies}. A recent work by \cite{Baes2020NonparametricWavelengths} explored the change in asymmetry in a range of wavelengths from the UV to the submm, as well as subsequent derived data products such as stellar mass, dust mass, and star-formation rate. Though they saw trends reflective of understood physical models, without assessing the varying signal-to-noise levels of those observations it is not clear how much of those trends are driven by noise effects.

\subsubsection{Example of effects of noise on merger asymmetries}

The changes in asymmetry from noise contribution could lead to observing trends that do not exist, but they could also result in obscuring relationships we expect to see. We can use our methods of adding noise to simulations, and correcting for that noise with different methods, to investigate the noise effects on an example science question: does galaxy asymmetry increase as the result of galaxy-galaxy interaction.

Observational studies have reliably found that photometric asymmetry increases for galaxy pairs as the separation between the interacting pairs decreases \citep{Hernandez-Toledo2005THEEVOLUTION,DePropris2007TheRate,Casteels2014GalaxyInformation}. Asymmetric structures resulting from an interaction are tied to the tidal forces disturbing the galaxy morphology, therefore the correlation between pair separation and asymmetry should be apparent in both optical light and stellar mass maps. This serves as a good test case for how the heightened uncertainty of a mass measurement would effect the prominence of this trend.

To test how noise will impact asymmetry measurements of TNG stellar mass maps, we collect a new sample of interacting TNG galaxies on which we can apply the four different asymmetry measurements we have discussed. We select a sample of TNG galaxies at z=0 with $10^9\leq \textrm{M}_{\star}/\textrm{M}_{\odot}$ which are undergoing a major merger, i.e. they have a companion with a mass ratio greater than 0.1 that is closer than 100 kpc in 3D separation. We select 300 interacting galaxies from TNG, with the goal of having approximately (though not always exactly) 30 galaxies in each 10 kpc bin for 3D separations ($r_{3D}$) ranging from 0-100kpc.

\begin{figure*}
	\includegraphics[width=0.75\textwidth]{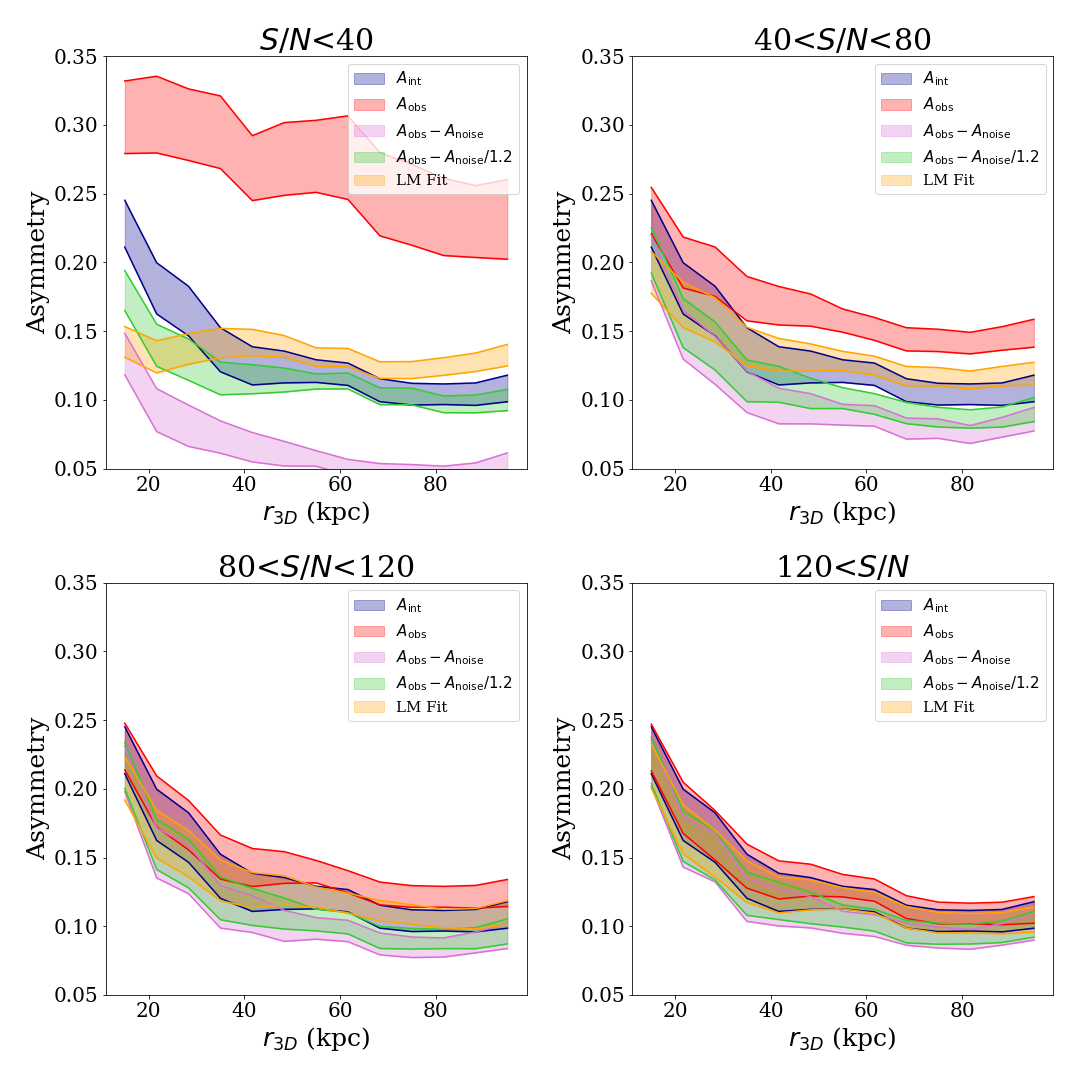}
	\centering
    \caption{The asymmetry of interacting galaxies in TNG as a function of their 3D separation from the nearest companion ($r_{3D}$). We include both the intrinsic asymmetry measurement (navy, identical in all panels), the observed asymmetry measurement (red), the observed asymmetry corrected for $A_{\rm noise}$ (pink), the observed asymmetry corrected for $A_{\rm noise}/1.2$ (green), and the asymmetry predicted by Equation \ref{eqn:A_noise_fit} (yellow). The four panels represent four different signal-to-noise ranges we apply to the same set of TNG galaxies, demonstrating how the trend between asymmetry and $r_{3D}$ is obscured in observed asymmetry as signal-to-noise decreases. Ideally all curves would match the navy curve, but as the $S/N$ regime decreases the different asymmetry measurements struggle to capture the true underlying trend between $A_{\rm int}$ and $r_{3D}$.}
    \label{fig:A_rp}
\end{figure*}

We then apply 4 different $S/N$ regimes to this data: $S/N<40$, $40<S/N<80$, $80<S/N<120$, and $S/N>120$, using the same methods described in Section \ref{sec:CAS}. Figure \ref{fig:A_rp} demonstrates how the relationship between asymmetry and pair separation varies between the intrinsic value (navy), the observed value (red), the observed corrected for $A_{\rm noise}$ (pink), the observed corrected for $A_{\rm noise}/1.2$ (green), and the asymmetry predicted from Equation \ref{eqn:A_noise_fit} (the LM-Fit method, yellow). If we only examine these changes for small amounts of added noise ($80<S/N<120$ and $S/N>120$), the observed asymmetry traces the intrinsic asymmetry's relationship to $r_{3D}$ almost perfectly. Once we enter the intermediate signal-to-noise regime ($40<S/N<80$), the trend has diminished for observed asymmetry. This is because the magnitude of $A_{\rm noise}$ depends on $A_{\rm obs}$. Thus, at low $r_{3D}$ values, where all galaxies have lower asymmetries compared to those in a later stage of interaction, the noise will have a greater impact on the observed asymmetry. As a result, galaxies at large separations will have an overestimated asymmetry, whereas the asymmetry of close galaxy pairs will be relatively unchanged by noise contribution. 

Given the change in asymmetry from noise differs with the intrinsic asymmetry, for the three lower signal-to-noise regimes in Figure \ref{fig:A_rp} some noise correction needs to be completed to capture the true relationship between asymmetry and pair separation. Below $S/N<120$ the best traditional noise correction to recover the asymmetry trend at high $r_{3D}$ is subtracting $A_{\rm noise}/1.2$ from $A_{\rm obs}$. One can also use the LM-Fit method to correct for noise and achieve similar recovery, and this works well in the intermediate $S/N$ regime. In the lowest signal-to-noise bin ($S/N<40$), the relationship between asymmetry and $r_{3D}$ has completely disappeared for the observed asymmetry. Correcting with the $A_{\rm noise}/1.2$ results in a trend closer to the intrinsic asymmetry, but it is still not as drastic a change in asymmetry between $r_{3D}$ values as expected. The LM-Fit method, though better replicating the intrinsic asymmetry than the traditional $A_{\rm noise}$ correction, also doesn't capture the complete trend between asymmetry and pair separation due to the large scatter in the fit for $S/N<50$. The degredation of this relationship with increased noise is an important example of how, even within a dataset of relatively uniform signal-to-noise values, the effects of $A_{\rm noise}$ can still alter the final conclusions. Those endeavouring to use non-parametric morphologies in their analysis will need to consider what degree variations (or lack-there-of) in asymmetry will result from noise effects, and what kind of noise correction is required for their desired analysis.

\section{Conclusions}
\label{sec:summary}

Using stellar mass maps from the IllustrisTNG100-1 and Illustris-1 simulations, we investigated the impact of observational noise and resolution, and the accuracy of different methods to recover the intrinsic asymmetry and concentration parameters. Our main conclusions are the following:

\begin{itemize}
  \item The traditional asymmetry metric is systematically biased by the presence of measurement uncertainties. The commonly used method of correcting for noise asymmetry contribution underestimates the asymmetry, particularly at $S/N<100$, where the asymmetry can be overestimated anywhere from a factor of 2 to 10 (see Figure \ref{fig:Int_Noise} and \ref{fig:Noise_Corr}).
  \item An alternative correction method, which subtracts $A_{\rm noise}/1.2$ from the observed asymmetry reduces this systematic bias, reaching less than 0.05 difference by $S/N\sim25$ (Figure \ref{fig:divide_by}).
  \item One can fit the relationship between the observed asymmetry and the signal-to-noise to predict the intrinsic asymmetry (see Equation \ref{eqn:A_noise_fit}), removing the bias in asymmetry created by noise as low as $S/N\sim5$ (Figure \ref{fig:A_PSF_corr_fit}).
  \item Degraded resolution also changes the asymmetry (and concentration) measurement from its intrinsic value. A fit relationship is determined between observed asymmetry or concentration and resolution (Res=$R_{\rm half}$/FWHM) to correct for this bias (see Equations \ref{eqn:A_fit_PSF} and \ref{eqn:C_fit}, respectively) and which accurately replicate the intrinsic values (Figures \ref{fig:A_PSF_corr_fit} and \ref{fig:C_PSF_fit}). Real asymmetries will need to be corrected for both resolution and noise, and though Equation \ref{eqn:A_piecewise} does not correct asymmetry as well as when just resolution needs to be accounted for (compare Figure \ref{fig:A_piecewise_correct} to Figure \ref{fig:A_PSF_corr_fit}), it still brings the majority of asymmetry measurements within $A_{\rm int}\pm0.05$.
  \item The noise and resolution correction work equally well on an unseen set of TNG galaxies (confirming the corrections are not overfit to their training data) and a set of Illustris galaxies (demonstrating these corrections can be applied to a simulation using different physical models). See Figure \ref{fig:Check_overfit_TNG} for details.
  \item We list various examples to demonstrate how the choice of asymmetry noise correction can alter science results, including our own hypothetical investigation into the relationship between asymmetry and galaxy interactions. Tables \ref{tab:A_noise_corr_TNG} provides an important summary of the biases created by different corrections for $S/N$ values to help future works ascertain the best method for their intended study. 
\end{itemize}

\section*{Acknowledgements}

The authors thank David Patton and Scott Wilkinson for their helpful discussion on the work herein, as well as the anonymous referee for their helpful report. MDT acknowledges the receipt of a Mitacs Globalink research award. AFLB \& RM gratefully acknowledge ERC grant 695671 'Quench' and support from the UK Science and Technology Facilities Council (STFC). MHH acknowledges support from William and Caroline Herschel Postdoctoral Fellowship Fund.

The simulations of the IllustrisTNG project used in this work were undertaken with compute time awarded by the Gauss Centre for Supercomputing (GCS) under GCS Large-Scale Projects GCS-ILLU and GCS-DWAR on the GCS share of the supercomputer Hazel Hen at the High Performance Computing Center Stuttgart (HLRS), as well as on the machines of the Max Planck Computing and Data Facility (MPCDF) in Garching, Germany.

\section*{Data Availability}
Simulation data from TNG100-1 and Illustris-1 used in this work is openly available on the IllustrisTNG website, at \hyperlink{tng-project.org/data}{tng-project.org/data}. 

%%%%%%%%%%%%%%%%%%%%%%%%%%%%%%%%%%%%%%%%%%%%%%%%%%


% \begin{thebibliography}
% \small
% \itemindent -0.48cm
    
% \TOCadd{Bibliography}
\bibliographystyle{mnras}
% \setcitestyle{authoryear,open={(},close={)}}
\bibliography{references.bib}

% \end{thebibliography}

% \TOCadd{Bibliography}
% \bibliographystyle{mnras}
% % 	\setcitestyle{authoryear,open={(},close={)}}
% \bibliography{Mendeley}
% %%%%%%%%%%%%%%%%%%%%%%%%%%%%%%%%%%%%%%%%%%%%%%%%%%

%%%%%%%%%%%%%%%%% APPENDICES %%%%%%%%%%%%%%%%%%%%%

% \appendix

% \section{Some extra material}

% If you want to present additional material which would interrupt the flow of the main paper,
% it can be placed in an Appendix which appears after the list of references.

%%%%%%%%%%%%%%%%%%%%%%%%%%%%%%%%%%%%%%%%%%%%%%%%%%

% Don't change these lines
\bsp	% typesetting comment
\label{lastpage}
\end{document}